\documentclass{svmult}

\usepackage{type1cm}        

\usepackage{amssymb}
\usepackage{makeidx}         
\usepackage{graphicx}        
\usepackage{multicol}        
\usepackage[bottom]{footmisc}

\makeindex             

\renewcommand{\greeksym}[1]{{\usefont{U}{psy}{m}{n}#1}}

\def\picSin{((\mathbf{p}_1 \times \mathbf{S}_1) \cdot \mathbf{n}_{1 2})}
\def\piicSin{((\mathbf{p}_2 \times \mathbf{S}_1) \cdot \mathbf{n}_{1 2})}
\def\piicSiin{((\mathbf{p}_2 \times \mathbf{S}_2) \cdot \mathbf{n}_{1 2})}
\def\picSiin{((\mathbf{p}_1 \times \mathbf{S}_2) \cdot \mathbf{n}_{1 2})}
\def\picSipii{((\mathbf{p}_1 \times \mathbf{S}_1) \cdot \mathbf{p}_2)}
\def\SiSii{(\mathbf{S}_1 \cdot \mathbf{S}_2)}
\def\pipii{(\mathbf{p}_1 \cdot \mathbf{p}_2)}
\def\Sipi{(\mathbf{S}_1 \cdot \mathbf{p}_1)}
\def\Siipii{(\mathbf{S}_2 \cdot \mathbf{p}_2)}
\def\Sipii{(\mathbf{S}_1 \cdot \mathbf{p}_2)}
\def\Siipi{(\mathbf{S}_2 \cdot \mathbf{p}_1)}
\def\Sin{(\mathbf{S}_1 \cdot \mathbf{n}_{1 2})}
\def\Siin{(\mathbf{S}_2 \cdot \mathbf{n}_{1 2})}
\def\pin{(\mathbf{p}_1 \cdot \mathbf{n}_{1 2})}
\def\piin{(\mathbf{p}_2 \cdot \mathbf{n}_{1 2})}
\def\pipi{\mathbf{p}_1^2}
\def\piipii{\mathbf{p}_2^2}

\begin{document}
\title*{Post-Newtonian methods: Analytic results on the binary problem}

\author{Gerhard Sch\"afer}

\institute{Gerhard Sch\"afer
	\at Friedrich-Schiller-Universit\"at Jena, Theoretisch-Physikalisches Institut,\\
    Max-Wien-Platz 1, D-07743 Jena, Germany\\
	\email{G.Schaefer@tpi.uni-jena.de}
}

\maketitle

\abstract*{A detailed account is given on approximation schemes to the Einstein theory of general relativity where the iteration starts from the Newton theory of gravity. Two different coordinate conditions are used to represent the Einstein field equations, the generalized isotropic ones of the canonical formalism of Arnowitt, Deser, and Misner and the harmonic ones of the Lorentz-covariant Fock-de Donder approach. Conserved quantities of isolated systems are identified and the Poincar\'e algebra is introduced. Post-Newtonian expansions are performed in the near and far (radiation) zones. The natural fitting of multipole expansions to post-Newtonian schemes is emphasized. The treated matter models are ideal fluids, pure point masses, and point masses with spin and mass-quadrupole moments modelling rotating black holes. Various Hamiltonians of spinning binaries are presented in explicit forms to higher post-Newtonian orders. The delicate use of black holes in post-Newtonian expansion calculations and of the Dirac delta function in general relativity find discussions.}

\abstract{A detailed account is given on approximation schemes to the Einstein theory of general relativity where the iteration starts from the Newton theory of gravity. Two different coordinate conditions are used to represent the Einstein field equations, the generalized isotropic ones of the canonical formalism of Arnowitt, Deser, and Misner and the harmonic ones of the Lorentz-covariant Fock-de Donder approach. Conserved quantities of isolated systems are identified and the Poincar\'e algebra is introduced. Post-Newtonian expansions are performed in the near and far (radiation) zones. The natural fitting of multipole expansions to post-Newtonian schemes is emphasized. The treated matter models are ideal fluids, pure point masses, and point masses with spin and mass-quadrupole moments modelling rotating black holes. Various Hamiltonians of spinning binaries are presented in explicit forms to higher post-Newtonian orders. The delicate use of black holes in post-Newtonian expansion calculations and of the Dirac delta function in general relativity find discussions.}

\section{Introduction}	\label{intro}
In the weak-field slow-motion limit of any theory of gravity the Newtonian theory of gravity comes into play because it describes the motion and structure of gravitating objects very well in that regime. According to current knowledge from experiments and observations, the most reliable theory of gravity is the Einstein theory of general relativity, \cite{will06}. 

Most of the objects in the Universe seem to have velocities $v$ which are small compared with the speed of light (in the following denoted by $c$), i.e. $v/c  \lesssim 1/3$.  In those cases it is to be expected that the Newtonian theory is a good starting point for an iteration scheme toward full general relativity thinking in terms of expansion of general relativity in powers of dimensionless $v/c$. The conservative nature of most of the phenomena, i.e. no measurable gravitational radiation damping, the power series has to be a one with ordering parameter $(v/c)^2$ (recall $v^2/c^2 \lesssim 1/10$ in the regime of our interest) because of motion-inversion symmetry in those cases. We shall see later that at  the order $(v/c)^5$ the gravitational radiation damping enters for the first time. Calling the order $(v/c)^{2n}$ the $n$th post-Newtonian order (n PN), the gravitational dissipation (because of gravitational radiation damping from radiation emission) enters at the 2.5 PN order so the exponent $n$ will also include half-integer numbers. In case of bound systems the virial theorem holds which tells us that  $v^2 \sim GM/r$, where $G$ is the Newtonian gravitational constant, $r$ a typical distance of two bodies or a typical radius of one body, and $M$ the total mass of the system. So an expansion of general relativity in powers of $(v/c)^2$ is at the same time an expansion in powers of  $GM/c^2r$ if bound systems are considered. From the dynamics of bound systems the dynamics of low-velocity scattering is straightforwardly obtained.

Depending on the mathematical representation of the Einstein theory (choice of coordinates, choice of variables, etc.) and on the physical aspects under investigation (physics in the near zone, physics in the far zone, etc.) there exist many of those expansions which all are called PN expansions. The crucial ordering parameter in all PN expansions is always $1/c^2$. The Einstein field equations as well as the equations of motion which follow from them by integrability conditions can be formally expanded in powers of $1/c^2$. The solutions of these equations, however, can not be expanded in this way, in general. Only under conditions where no gravitational radiation is present an all-over-in-space expansion in powers of $1/c^2$ is feasible. This is the case for stationary systems only. Yet for those parts of a radiating system where radiation plays no r{\^o}le, i.e. in the conservative parts, an expansion in powers of $1/c^2$ can be carried out too. In general, PN expansions in powers of $1/c$ are feasible in the near zone ($r << \lambda$ with $\lambda$ a typical wavelength of the gravitational radiation and $r$ the radius of a sphere enclosing the matter source) and in the far or radiation zone ($r >> \lambda$) of a radiating system but they are not analytic in $1/c$ because $c$ is showing up in log-terms. In the near zone, up to 3.5 PN order, which means order $1/c^7 = (1/c^2)^{3.5}$, a PN expansion in powers of $1/c$ is valid, where the even powers are connected with the conservative dynamics and the odd powers with the dissipative radiation reaction dynamics. From the 4 PN order on (corresponding to 1.5 PN order in the far zone), log-$c$ terms enter via tail effects which emerge from back scattering, \cite{blanchet06}. Post-Minkowskian (PM) series are global series in space where only the weak-field limit is assumed without restrictions to the velocities of the bodies. Here, the ordering parameter is $G$ or, dimensionless $GM/c^2r$, which is equivalent to a non-linearity expansion of the Einstein field equations, \cite{blanchet06}. If the virial theorem holds, namely $(GM/c^2r) \sim (v/c)^2$, PM series can be further expanded into PN series. Having much less information on their respective orders, the PN series are much easier to be worked out analytically, compare e.g., the 1 PN point-mass Hamiltonian, see e.g. \cite{landau85}, with the 1 PM point-mass Hamiltonian, only recently achieved in \cite{ledvinka08}.

The most extensively performed PN calculations in the literature can be divided into two classes, the class applying harmonic coordinate conditions, see e.g. \cite{fock66}, and the other class using Arnowitt-Deser-Misner (ADM) generalized isotropic coordinate conditions \cite{ADM62}. Whenever detailed comparisons between the two classes have been made, agreement of the results could be achieved, see e.g. \cite{damour0102}, \cite{blanchet03}. This is quite an important aspect in view of the use of Dirac delta functions for modelling black holes in explicit calculations. These functions are a-priori problematic in a non-linear theory like general relativity so sophisticated regularization methods have to be employed. The only regularization method which turned out to be successful to the highest orders of the explicit calculations is the dimensional one, \cite{damour01}, \cite{blanchet04a}. In sect. \ref{sec2_3} a detailed account of black holes together with their representing Dirac delta functions will be given because they blatantly violate the weak-field condition. In this case, only external velocities of the black holes can be small and their relative gravitational interaction weak, and only in this sense black holes and Dirac delta functions can fit into PN schemes.

The binary point-mass dynamics at 3 PN order and the gravitational waves emitted hereof (3 PN wave generation) could have been managed successfully through only dimensional regularization, \cite{damour01}, \cite{blanchet04a}, \cite{blanchet04b}. Extended body calculations have been performed too but they are much more complicated and still not fully under control at 3 PN order \cite{pati02}. Only a surface-integral-based method of the Einstein-Infeld-Hoffmann type did succeed too, \cite{itoh04}, \cite{futamase07}.

The many-body systems treated most successfully with PN approximation methods are point-mass systems composed from two or more objects and spinning binary systems with even rotationally deformed components; for the most recent results with spin, see \cite{faye06}, \cite{blanchet0607}, \cite{damour08}, \cite{hergt08a}, \cite{hergt08b}, \cite{steinhoff08a}, \cite{steinhoff08b}.

Recently, an effective field theory approach has been advocated for PN calculations \cite{goldberger06}. Applications to point-mass systems and to systems with spinning components have already been performed through higher PN orders, \cite{porto08a}, \cite{porto08b}, \cite{gilmore08}. An obvious difference in the spin dynamics between \cite{steinhoff08b} and \cite{porto08b} found clarification in \cite{steinhoff09} showing the correctness of the result in \cite{steinhoff08b}.

In this article, Latin indices from the mid alphabet are mostly running from 1 to 3, ${\bf v} = (v^i), {\bf v}^2 = v^iv^i$, $\partial_t$ and $\partial_i$ denote the partial derivatives with respect to the time and space coordinates $t$ and $x^i$, respectively, and the functional derivative (Fr\'echet derivative) of a functional, say $F[f]$, with respect to a function $f$ takes the form $\delta F[f]/ \delta f$.  Greek indices are mostly running from 0 to 3, whereby $x^0 = ct$. The signature of the four-dimensional metric $g_{\mu\nu}$ is +2. Particles are numbered with indices from the beginning of the Latin alphabet.

\section{Systems in Newtonian gravity in canonical form}	\label{sec1}
In the Newtonian theory, the equations governing the motion of gravitating ideal fluids are (i) the equation for the conservation of mass, \begin{equation}
\partial_t \varrho_* + \mbox{div}( \varrho_* {\bf v}) = 0\,,
\end{equation}
where $\varrho_*$ is the mass density and ${\bf v}$ the velocity fied of the fluid, (ii) the equations of motion,
\begin{equation}
\varrho_* \partial_t {\bf v} + \frac{\varrho_*}{2}~ \mbox{grad}~{\bf
v}^2 -  \varrho_*~ {\bf v} \times  \mbox{curl}~  {\bf v}  = - \mbox{grad}~p  + \varrho_*~\mbox{grad}~U\,,
\end{equation}
where $p$ is the pressure in the fluid and $U$ the gravitational potential, and (iii) the equation of state, using internal energy density $\epsilon$ and specific enthalpy $h$,
\begin{equation}
\epsilon = \epsilon(\varrho_*, s) \quad \mbox{with} \quad d\epsilon = h
d\varrho_*+\varrho_*Tds,  \quad \mbox{or} \quad dp = \varrho_*dh - \varrho_*Tds \,,
\end{equation}
with (iv) the conservation law for the specific entropy $s$ along the flow lines,
\begin{equation}
\partial_t s + {\bf v}\cdot  \mbox{grad}~s = 0\,.
\end{equation}
The gravitational potential reads,
\begin{equation}
U({\bf x},t) = G \int d^3{\bf x}' \frac{ \varrho_*({\bf x}',t)}{|{\bf x}-{\bf x}'|}\,.
\end{equation}
It results from (v) the Newtonian field equation,
\begin{equation}
\Delta U = - 4\pi G \varrho_*\,.
\end{equation}
$\Delta$ is the Laplacian and $|...|$ means the standard euclidean distance.

Written in form of Hamilton equations of motion, i.e. $\partial_t A({\bf x},t) = \{A({\bf x},t), H\}$, the equations above take the forms, (i) the mass conservation equation
\begin{equation}
\frac{\partial \varrho_*}{\partial t} = - \partial_i \left(\frac{\delta H}{\delta \pi_i}\varrho_*\right)\,,
\end{equation}
notice $ v^i = \frac{\delta H}{\delta \pi_i}$, (ii) the equations of motion
\begin{equation}
\frac{\partial \pi_i}{\partial t} = - \partial_j \left(\frac{\delta H}{\delta \pi_j}\pi_i\right)
 - \partial_i \left(\frac{\delta H}{\delta \pi_j}\right) \pi_j - \partial_i \left(\frac{\delta
H}{\delta \varrho_*}\right)\varrho_* + \frac{\delta H}{\delta s}  \partial_i s\,,
\end{equation}
and (iv) the entropy conservation law
\begin{equation}
\frac{\partial s}{\partial t} = - \frac{\delta H}{\delta \pi_i}  \partial_i s\,,
\end{equation}
where the Hamiltonian is given by $H = H(\varrho_*, \pi_i, s)$ with $\pi_i$ the linear momentum density of the fluid, see \cite{holm85}. Here, use has been made of the kinematical Lie-Poisson bracket relations between the fundamental variables
\begin{equation}
\{\pi_i({\bf x},t),\varrho_*({\bf x}',t)\}  = \frac{\partial}{\partial x'^i}[\varrho_*({\bf x}',t) \delta ({\bf x}-{\bf x}')]\,,	\label{eq10}
\end{equation}
\begin{equation}
\{\pi_i({\bf x},t),s({\bf x}',t)\}  = \frac{\partial s({\bf x}',t)}{\partial {x}'^i} \delta ({\bf x}-{\bf x}')\,,
\end{equation}
\begin{equation}
\{\pi_i({\bf x},t),\pi_j({\bf x}',t)\}  = \pi_i({\bf x}',t)\frac{\partial}{\partial {x}'^j} \delta ({\bf x}-{\bf x}')
- \pi_j({\bf x},t)\frac{\partial}{\partial {x}^i} \delta ({\bf x}-{\bf x}')\,,	\label{eq12}
\end{equation}
and zero otherwise, where $ \delta ({\bf x}-{\bf x}')$ denotes the standard Dirac delta function in three-dimensional space. It fulfills $\int d^3x ~\delta ({\bf x})= 1$.

In the Newtonian theory, the Hamiltonian of the fluid is given by,
\begin{equation}
H = \frac{1}{2}\int d^3{\bf x} \frac{\pi_i\pi_i}{\varrho_*} - \frac{G}{2}\int d^3{\bf x}d^3{\bf x}'
\frac{\varrho_*({\bf x},t)\varrho_*({\bf x}',t)}{|{\bf x}-{\bf x}'|} + \int d^3{\bf x}~ \epsilon\,.
\end{equation}

For point masses, the total momentum and mass densities read, consistent with the Eqs. (\ref{eq10}) and (\ref{eq12}),
  \begin{equation}
\pi_i = \sum_a p_{ai} \delta ({\bf x}-{\bf x}_a), \quad\quad \varrho_* = \sum_a m_a \delta ({\bf x}-{\bf x}_{a})\,,
\end{equation}
where the position and momentum variables fulfill the standard Poisson bracket relations,
\begin{equation}
\{x^i_a,p_{aj}\} = \delta_{ij}, \quad\quad \mbox{zero otherwise}\,,
\end{equation}
and the Hamiltonian takes the form,
\begin{equation}
H = \frac{1}{2} \sum_a \frac{p_{ai}p_{ai}}{m_a} - \frac{G}{2}  \sum_{a \ne b} \frac{m_am_b}{|{\bf x}_a-{\bf x}_b|}\,,
\end{equation}
where the self-energy term has been dropped (for regularization techniques, see section \ref{sec2_3}).

\section{Canonical general relativity and PN expansions}	\label{sec2}
In curved spacetime the stress-energy tensor of an ideal fluid takes the form
\begin{equation}
T^{\mu\nu} = \varrho(c^2 + h) u^{\mu}u^{\nu} + p g^{\mu\nu}, \quad\quad g_{\mu\nu}u^{\mu}u^{\nu} = -1\,,
\end{equation}
where $\varrho$ denotes the proper rest-mass density, $h$ the specific enthalpy, and $u^{\mu}$ the four-velocity field of the fluid. Using energy density $e= \varrho(c^2 + h) - p$ (also the specific internal energy $\Pi = e/\varrho -c^2$ could be used), the equation of state reads
\begin{equation}
e = e(\varrho,s) \quad \mbox{with} \quad de=(c^2+h)d\varrho + \varrho Tds, \quad  \mbox{or}  \quad dp = \varrho dh - \varrho Tds\,.
\end{equation}
The variables of the canonical formalism are chosen to be
\begin{equation}
\varrho_* = \sqrt{-g} u^0 \varrho, \quad\quad s, \quad\quad \pi_i = \frac{1}{c} \sqrt{-g} T^0_i.
\end{equation}
They fulfill the same (universal) kinematical Lie-Poisson bracket relations as in the Newtonian theory, see \cite{holm85}, or also \cite{blanchet90},
\begin{equation}
\{\pi_i({\bf x},t),\varrho_*({\bf x}',t)\}  = \frac{\partial}{\partial x'^i}[\varrho_*({\bf x}',t) \delta ({\bf x}-{\bf x}')]\,,
\end{equation}
\begin{equation}
\{\pi_i({\bf x},t),s({\bf x}',t)\}  = \frac{\partial s({\bf x}',t)}{\partial x'^i} \delta ({\bf x}-{\bf x}')\,,
\end{equation}
\begin{equation}
\{\pi_i({\bf x},t),\pi_j({\bf x}',t)\}  = \pi_i({\bf x}',t)\frac{\partial}{\partial x'^j} \delta ({\bf x}-{\bf x}')
- \pi_j({\bf x},t)\frac{\partial}{\partial x^i} \delta ({\bf x}-{\bf x}')\,.
\end{equation}
The evolution equations take the form
\begin{equation}
\frac{\partial \varrho_*}{\partial t} = - \partial_i \left(\frac{\delta
H}{\delta \pi_i}\varrho_*\right) \quad \Longleftrightarrow \quad \partial_{\mu}(\sqrt{-g}\varrho u^{\mu})=0\,,
\end{equation}
\begin{equation}
\frac{\partial s}{\partial t} = - \frac{\delta
H}{\delta \pi_i}  \partial_i s \quad \Longleftrightarrow \quad {u^{\mu}\partial_{\mu}s=0}\,,
\end{equation}

\begin{eqnarray}
\frac{\partial \pi_i}{\partial t} = - \partial_j \left(\frac{\delta H}{\delta \pi_j}\pi_i\right)
 - \partial_i \left(\frac{\delta H}{\delta \pi_j}\right)\pi_j - \partial_i \left(\frac{\delta
H}{\delta \varrho_*}\right)\varrho_* &+& \frac{\delta H}{\delta s}  \partial_i s  \\ [2ex]\nonumber
\mbox{corresponding to}
\quad \partial_{\mu}\left(\sqrt{-g}\,T^{\mu}_i\right)
- \frac{1}{2}\, \sqrt{-g}\,T^{\mu\nu}\, \partial_i g_{\mu\nu} &=& 0 \,,
\end{eqnarray}
\begin{equation}
v^i = \frac{\delta H}{\delta \pi_i}, \quad \mbox{where} \quad v^i = c \frac{u^i}{u^0}\,.
\end{equation}

The linear and angular momenta of the fluid read, respectively,
\begin{equation}
P_i = \int d^3{\bf x} ~\pi_i, \quad\quad J_i = \int d^3{\bf x}~ \epsilon_{ijk}x^j\pi_k\,.
\end{equation}

For a system made of point masses simplifications take place,
\begin{equation}
h=p=s=0\,, \quad\quad (\mbox{dusty matter}),
\end{equation}
and further,
\begin{equation}
\varrho_* = \sum_a m_a  \delta ({\bf x}-{\bf x}_{a}), \quad\quad
\pi_i = \sum_a p_{ai} \delta ({\bf x}-{\bf x}_a), \quad\quad v^i_a=\frac{dx^i_a}{dt},
\end{equation}
where $p_{ai}$ and $x^i_a$ respectively are the linear momentum and the position vector of the $a$th particle. The kinematical Poisson bracket relations are given by
\begin{equation}
\{x^i_a,p_{aj}\} = \delta_{ij}, \quad\quad \mbox{zero otherwise}\,.
\end{equation}
Hereof the standard Hamilton equations result,
\begin{equation}
\frac{dp_{ai}}{dt} =  - \frac{\partial H}{\partial x^i_a}\,, \qquad \frac{dx^i_a}{dt} = \frac{\partial H}{\partial p_{ai}}\,.
\end{equation}
Remarkably, the difference to the Newtonian theory comes solely from the Hamiltonian which is thus a dynamical difference and not a kinematical one. This statement refers to the matter only and not to the gravitational field. The latter is quite different in general relativity. 

\subsection{Canonical variables of the gravitational field }	

Within the ADM canonical formalism of general relativity, in generalized isotropic coordinates, the independent gravitational field variables $h^{\rm TT}_{ ij}$ and $\pi_{\rm TT}^{ij}$ enter in the form
\begin{eqnarray}
g_{ij} =\left(1+\frac{1}{8}\phi\right)^4 \delta_{ij}+h^{\rm TT}_{ ij}\,,								\label{eq32}
\end{eqnarray}
\begin{eqnarray}
\pi^{ij}={\tilde{\pi}^{ij}}+\pi_{\rm TT}^{ij}\,,\label{eq33}
\end{eqnarray}
where $g_{ij}=g_{ji} \equiv \gamma_{ij}$ is the metric of the curved three-dimensional hypersurfaces $t$ = const, $\pi^{ij}c^3/16\pi G$ is the canonical conjugate to $\gamma_{ij}$, i.e. $\pi^{ij} = - \gamma^{1/2} (K^{ij} - \gamma^{ij} K^k_k)$, where $K_{ij}=K_{ji}$ is the extrinsic curvature of the $t$ = const slices, and $\gamma = \mbox{det}(\gamma_{ij})$,  $\gamma^{il} \gamma_{lj}= \delta_{ij}$, and for $\tilde{\pi}^{ij}$ holds
\begin{eqnarray}
\tilde{\pi}^{ij}=\partial_i \pi^j + \partial_j \pi^i - \frac{2}{3}\delta_{ij} \partial_k \pi^k.			\label{eq34}
\end{eqnarray}
Obviously, $\pi^{ii}=0$, or $\pi^i_i = \pi^{ij} h^{\rm TT}_{ij}$. The canonical conjugate to $h^{\rm TT}_{ij}$ reads $\pi_{\rm TT}^{ij}c^3/16\pi G$. The index ${\rm TT}$ means tranverse-traceless, i.e. $h^{\rm TT}_{ii} = \pi_{\rm TT}^{ii} = 0$, $\partial_j h^{\rm TT}_{ij} = \partial_j \pi_{\rm TT}^{ij}=0$.

Using those variables, the Einstein field equation $\sqrt{-g}G^{00} = \frac{8\pi G}{c^4} \sqrt{-g}T^{00}$ can be put into the form, employing point masses for the source,
\begin{eqnarray}
\gamma^{1/2}\mbox{R} = \frac{1}{\gamma^{1/2}}\left(\pi^i_j \pi^j_i-\frac{1}{2}\pi^i_i\pi^j_j \right)
+\frac{16\pi G}{c^3} \sum_a\left(m_a^2c^2 + \gamma^{ij}p_{ai}p_{aj}\right)^{1/2}\delta_a\,,									\label{eq35}
\end{eqnarray}
and the field equations $\sqrt{-g}G^{0}_i = \frac{8\pi G}{c^4} \sqrt{-g}T^{0}_i$ read
\begin{eqnarray}
-2 \partial_j \pi^j_i + \pi^{kl} \partial_i \gamma_{kl}  = \frac{16\pi G}{c^3} \sum_a p_{ai}\delta_a\,,							\label{eq36}
\end{eqnarray}
where $\delta_a = \delta ({\bf x}-{\bf x}_a)$. The Eqs. (\ref{eq35}) and (\ref{eq36})  are the famous four constraint equations of general relativity.

In the gauge Eqs. (\ref{eq32}) - (\ref{eq34}) the ADM Hamiltonian can be written, \cite{ADM62},
\begin{eqnarray}
H\left[x^i_a, p_{ai}, h^{\rm TT}_{ij}, \pi_{\rm TT}^{ij}\right] = - \frac{c^4}{16\pi G}\int d^3{\bf x}~
\Delta\phi\left[x^i_a, p_{ai}, h^{\rm TT}_{ij}, \pi_{\rm TT}^{ij}\right]\,,
\end{eqnarray}
resulting from the solution of the four (elliptic-type) constraint equations. The additional Hamilton equations of motion for the gravitational field are given by
\begin{equation}
\frac{\partial \pi_{\rm TT}^{ij}}{\partial t} =  - \frac{16\pi G}{c^3} \frac{\delta H}{\delta  h^{\rm TT}_{ij}}, \quad\quad
\frac{\partial h^{\rm TT}_{ij}}{\partial t} =   \frac{16\pi G}{c^3} \frac{\delta H}{\delta  \pi_{\rm TT}^{ij}}\,.
\end{equation}

The transition to a Routh functional simplifies a lot the construction of the dynamics of the matter and of the gravitational field. The Routh functional is chosen in the form, \cite{jaranowski98},
\begin{eqnarray}
R\left[x^i_a, p_{ai}, h^{\rm TT}_{ij}, \partial_th^{\rm TT}_{ij}\right]
= H - \frac{c^3}{16\pi G} \int d^3{\bf x} ~\pi^{ij}_{\rm TT} \partial_th^{\rm TT}_{ij}\,.
\end{eqnarray}
The evolution equations for the matter and the gravitational field now read
\begin{eqnarray}
\frac{\delta \int R(t')  dt'}{\delta h^{\rm TT}_{ij}(x^k,t)} = 0\,, \quad\quad
\dot{p}_{ai} = - \frac{\partial R}{\partial x^i_a}\,, \qquad \dot{x}^i_a = \frac{\partial R}{\partial p_{ai}}\,.
\end{eqnarray}
The conservative dynamics results from the on-field-shell Routh functional
\begin{eqnarray}
R_{\rm shell}(t) = R\left[x^i_a, p_{ai}, h^{\rm TT}_{ij}[x^k_a, p_{ak}], \partial_th^{\rm TT}_{ij}[x^k_a, p_{ak}]\right]\,,
\end{eqnarray}
with solved field equations, in the form
\begin{eqnarray}
\dot{p}_{ai}(t) = - \frac{\delta \int R_{\rm shell}(t') dt'}{\delta x^i_a(t)}, \quad \quad
\dot{x}^i_a(t) = \frac{\delta \int R_{\rm shell}(t') dt'}{\delta p_{ai}(t)}\,,
\end{eqnarray}
where
\begin{eqnarray}
\frac{\delta \int R_{\rm shell}(t') dt'}{\delta z(t)} = \frac{\partial R_{\rm shell}}{\partial z(t)}
- \frac{d}{dt} \frac{\partial R_{\rm shell}}{\partial \dot{z}(t)} + ...\,, \quad z = (x^i_a, p_{ai})\,.
\end{eqnarray}
Using the matter equations of motion in the Routhian $R_{\rm shell}$  the Routhian can be brought into the form  $R(x^i_a, p_{ai})$. Herein, however, the meaning of the variables $x^i_a$ and $p_{ai}$ has changed, see \cite{schaefer84}, \cite{damour91}, \cite{damja00}.

\subsection{Brill-Lindquist initial-value solution for binary black holes}	\label{sec2_1}
The Brill-Lindquist solution for multiple black holes is a pure vacuum solution of the constraint equations at initial time $t$ under the conditions of time symmetry, i.e. $p_{ai}=0=\pi^{ij}$, and of conformal flatness, i.e. $h^{\rm TT}_{ij} = 0$, \cite{BL63}. A related vacuum solution is the Misner-Lindquist solution where an additional isometry condition is imposed, \cite{M63}, \cite{L63}. Under those conditions, the only remaining constraint equation reads, not using vacuum but (point-mass) sources,
\begin{eqnarray}
- \left(1+\frac{1}{8}~\phi\right) \Delta \phi =
\frac{16\pi G}{c^2} \sum_a m_a \delta_a\,, \quad \quad  (h^{\rm TT}_{ij} = 0 = p_{ai}=\pi^{ij})\,.
\end{eqnarray}
In the case of two black holes, its solution is given by, see \cite{jaranowski99},
\begin{eqnarray}
\phi = \frac{4G}{c^2} \left(\frac{\alpha_1}{r_1} + \frac{\alpha_2}{r_2}\right)
\end{eqnarray}
with ($a,b = 1,2$ and $b \ne a$)
\begin{eqnarray}
\alpha_a = \frac{m_a-m_b}{2} + \frac{c^2r_{ab}}{G}\left(\sqrt{1+\frac{m_a+m_b}{c^2r_{ab}/G} +
\left(\frac{m_a-m_b}{2c^2r_{ab}/G}\right)^2} - 1 \right)\,,						\label{eq46}
\end{eqnarray}
resulting into the Brill-Lindquist solution for binary black holes. Obviously, each Brill-Lindquist black hole is represented by a Dirac delta function (fictitious image mass-point; see section \ref{sec2_3}). In the Misner-Lindquist case, infinite many fictitious image mass-points are needed for each black hole, \cite{M63}, \cite{L63}, \cite{jaranowski99}.

The energy of the Brill-Lindquist solution simply reads
\begin{eqnarray}
H_{\rm BL}= (\alpha_1 + \alpha_2)c^2 =(m_1 + m_2)c^2 - G~ \frac{\alpha_1\alpha_2}{r_{12}}\,,\label{eq47}
\end{eqnarray}
The methods which have been used for the obtention of the Brill-Lindquist solution from sources (notice,
in the original work of Brill and Lindquist this solution has been obtained without any regularization 
as a purely vacuum solution) are analytical Hadamard regularization and mass renormalization, \cite{jaranowski99}, as well as dimensional regularization based on the d-dimensional metric
\begin{eqnarray}
 \gamma_{ij}=\left(1+\frac{1}{4}\frac{d-2}{d-1}\phi\right)^{\frac{4}{d-2}} \delta_{ij}
\end{eqnarray}
with solution ($\Gamma$ denotes the Euler gamma function)
\begin{eqnarray}
\phi = \frac{4G}{c^2} \frac{\Gamma(\frac{d-2}{2})}{\pi^{\frac{d-2}{2}}}
           \left(\frac{\alpha_1}{r_1^{d-2}} + \frac{\alpha_2}{r_2^{d-2}}\right)
\end{eqnarray}
(for more details see section \ref{sec2_3}). The PN expansion of the Brill-Lindquist initial energy expression is straightforward to all orders of $1/c^2$. Once it has fixed the static ambiguity parameter $\omega_{static}$ (see \cite{damja00}) in non-dimensional-regularization calculations to the correct value of zero, \cite{jaranowski99}. At that time, however, it was not quite clear that the Brill-Lindquist solution delivers the correct boundary conditions for 
the point-mass model. 

The truncation of the constraint equations in the form $h^{\rm TT}_{ij} \equiv 0$ as well as dropping an additional term in the Hamiltonian constraint connected with the energy density of the field momentum results in a remarkable, fully explicitly solvable conservative so-called skeleton dynamics which allows a PN expansion of the Hamiltonian, and of all the metric coefficients too, to all orders, see next section.

\subsection{Skeleton Hamiltonian}	\label{sec2_2}
In Ref. \cite{faye04} the skeleton dynamics has been developed. The skeleton approach to general relativity requires the conformal flat condition for the spatial three-metric for all times (not only initially as for the Brill-Lindquist solution)
\begin{eqnarray}
\gamma_{ij}& = (1+\frac{1}{8}\phi)^4 \delta_{ij}\,.
\end{eqnarray}
Hereof, in our coordinate system, maximal slicing follows,
\begin{eqnarray}
\pi^{ij}\gamma_{ij} & = 2 \sqrt{\gamma} \gamma^{ij} K_{ij} = 0\,.
\end{eqnarray}
Under the conformal flat condition for the spatial three-metric, the momentum constraint equations become
\begin{equation}
\pi^{j}_{i,\,j} = - \frac{8 \pi G}{c^3} \sum_a p_{ai}\delta_a\,.
\end{equation}
The solution of these equations is constructed under the condition that $\pi^{j}_{i}$ (and not $\pi^{ij}$, see Eqs. (\ref{eq33}) and (\ref{eq34})) is purely longitudinal, i.e.
\begin{equation}
\pi^{j}_{i} = \partial_i V_j +  \partial_j V_i - \frac{2}{3} \delta_{ij} \partial_l V_l\,.
\end{equation}
This condition is part of the definition of the skeleton model. At spacelike infinity, the surface-area integrals of $\pi^{j}_{i}$ or $\pi^{ij}$ are proportional to the total linear momentum of the binary system.

Furthermore, in the Hamilton constraint equation, which in our case reads
\begin{eqnarray}
\Delta\phi = - \frac{ \pi^j_i \pi^i_j }{(1+\frac{1}{8}\phi)^{7} } -\frac{16\pi G}{c^2}\sum_a \frac{ m_a\delta_a}{ (1+\frac{1}{8}\phi)}\,
\biggl ( 1+ \frac{p_a^2}{(1+\frac{1}{8}\phi)^4m_a^2c^2} \biggr )^{1/2}\,,
\end{eqnarray}
we perform a truncation of the numerator of the first term in the following way
\begin{eqnarray}
\pi^j_i \pi^i_j \equiv -2 V_j\partial_i\pi^i_j + \partial_i(2V_j\pi^i_j) \,
 \rightarrow -2 V_j\partial_i\pi^i_j=\frac{16\pi G}{c^3}\sum_a p_{aj}V_j\delta_a\,,
\end{eqnarray}
i.e. we drop from $\pi^j_i \pi^i_j$ the term $\partial_i(2V_j\pi^i_j)$. This is the second crucial truncation condition additional to the conformal flat one. Without this truncation neither an explicit solution can be achieved nor a PN expansion is feasible. From \cite{jaranowski98} we know that at the 3 PN level the $h^{TT}_{ij}$-field is needed to make the sum of the corresponding terms from $\pi^j_i \pi^i_j$ analytic in $1/c$.

With the aid of the ansatz
\begin{eqnarray}
\phi = \frac{4G}{c^2}\sum_a \frac{\alpha_a}{r_a}
\end{eqnarray}
and by making use of dimensional regularization, the energy and momentum constraint equations result in an algebraic equation of the form, \cite{faye04},
\begin{eqnarray}
\alpha_a = \frac{m_a}{1+ A\,\frac{\alpha_b}{r_{ab}}} \left[1 + \frac{p_a^2/(m_a^2 c^2)}{\left(1+ A
\alpha_b/r_{ab} \right)^{4}} \right]^\frac{1}{2} + \frac{p_{ai} V_{ai}/c}{\left(1+ A \, \alpha_b/r_{ab} \right)^{7}}\,,	\label{eq57}
\end{eqnarray}
where $A \equiv G/(2\,c^2)$ and $b \ne a$.

With these inputs the skeleton Hamiltonian for binary black holes becomes (at least initially, for $p_a=0$, the solution is consistent with general relativity)
\begin{eqnarray}
H_{\rm sk} \equiv  - \frac{c^4}{16\pi G}\int d^3{\bf x}\,\Delta\phi = c^2 \sum_a \alpha_a\,.
\end{eqnarray}
The Hamilton equations of motion read
\begin{eqnarray}
\dot{ \bf x}_a = \frac{\partial H}{\partial {\bf p}_a}\,, \qquad \dot{\bf p}_{a} = - \frac{\partial H}{\partial {\bf x}_a} \,.
\end{eqnarray}
In the center-of-mass frame of the binary system, we define
\begin{eqnarray}
 \mathbf{p} \equiv {\bf p}_1 = - {\bf p}_2, \quad \mathbf{r}  \equiv \mathbf{x}_1- \mathbf{x}_2, \quad
r^2 = \left ( \mathbf{x}_1- \mathbf{x}_2 \right) \cdot \left ( \mathbf{x}_1- \mathbf{x}_2 \right )\,.				\label{eq60}
\end{eqnarray}
Further, we will employ the following convenient dimensionless quantities
\begin{eqnarray}
\hat t &=& \frac{t\, c^3}{G\,m}\,,\,\,\, \hat r = \frac{r\, c^2}{G\,m}\,,\,\,\, \mathbf{\hat p} = \frac{\mathbf{p}}{\mu\,c}\,,
\,\,\, {\hat H}_{\rm sk} = \frac{H_{\rm sk}}{\mu\,c^2}\,,\\ \hat j & =& \frac{J\,c}{G\,m\, \mu}\,,
\,\,\, \hat p_r = \frac{{p_r}}{\mu\,c}\,, \,\,\, \mathbf{\hat p}^2 = \hat p_r^2 + \hat j^2/\hat r^2\,,				\label{eq62}
\end{eqnarray}
where $\mathbf{J} = \mathbf{r} \times \mathbf{p}$ is the orbital angular momentum in the center-of-mass frame and $p_r = \mathbf{p} \cdot \mathbf{r}/r$ the radial momentum. The total rest-mass is denoted by $m=m_1+m_2$ and the reduced mass by $\mu=m_1m_2/m$. The binary skeleton Hamiltonian ${\hat H}_{\rm Sk}$ can be put into the following form, \cite{gopu08},
\begin{eqnarray}
{\hat H}_{\rm sk} &=& {2\, \hat r} \biggl ( \psi_1 + \psi_2 - 2 \biggr ) \quad {\rm with}\\[1ex] \psi_{1} &=&
1+\frac{ \chi_{-} }{4\, \hat r\, \psi_2} \, \biggl (1+ \frac{ 4\,{\nu}^{2}
\left( {{\hat p_r}}^{2}+{\hat j}^2/{\hat r}^{2} \right) }{ \chi_{-}^2\, \psi_2^4} \biggr )^{1/2}
- \frac{ \left( 8\,{{ \hat p_r}}^{2}+7{\hat j}^2/{\hat r}^2 \right)
{\nu}^{2}}{8\, {\hat r}^{2}\psi_2^7 } \,,  \label{eq64}             \\ [1ex] \psi_{2}            &=&
1+\frac{ \chi_{+} }{4\, \hat r\, \psi_1} \, \biggl (1+ \frac{ 4\,{\nu}^{2}
\left( {{\hat p_r}}^{2}+{\hat j}^2/{\hat r}^2 \right) }{ \chi_{+}^2\, \psi_1^4} \biggr )^{1/2} - \frac{
\left( 8\,{{ \hat p_r}}^{2}+7{\hat j}^2/{\hat r}^2 \right) {\nu}^{2}}{8\, {\hat r}^{2}\psi_1^7} \,,\label{eq65}
\end{eqnarray}
where $ \chi_{-} = \left( 1-\sqrt {1-4\,\nu} \right) $ and $\chi_{+} = \left( 1+\sqrt {1-4\,\nu} \right)$ with $\nu = \mu/m$.

The conservative skeleton Hamiltonian has the following nice properties. It is exact in the test-body limit where it describes the motion of a test particle in the Schwarzschild spacetime. It is identical to the 1 PN accurate Hamiltonian for the binary dynamics in general relativity. Further, as explained earlier, when point particles are at rest, the Brill-Lindquist initial value solution is reproduced. It is remarkable that the skeleton Hamiltonian allows a PN expansion in powers of $1/c^2$ to arbitrary orders. The skeleton Hamiltonian thus describes the evolution of a kind of black holes under both conformal flat conditions for the three-metric and analyticity conditions in $1/c^2$ for the Hamiltonian. Of course, gravitational radiation emission is not included. It can, however, be added to some reasonable extent, see \cite{gopu08}.

Restricting to circular orbits and defining $x= (Gm\omega/c^3)^{2/3}$, where $\omega$ is the orbital angular frequency, the skeleton Hamiltonian reads explicitly to 3 PN order,
\begin{eqnarray}
{\hat H}_{\rm sk} &= & -\frac{x}{2} + \bigg(\frac{3}{8}+\frac{\nu}{24}\bigg) x^2
+ \bigg(\frac{27}{16}+\frac{29}{16}\nu-\frac{17}{48}\nu^2\bigg) x^3 \nonumber\\[1ex]
& + &\bigg(\frac{675}{128}+\frac{8585}{384}\nu-\frac{7985}{192}\nu^2
+\frac{1115}{10368}\nu^3\bigg) x^4 + \mathcal{O}(x^{5})\,.
\end{eqnarray}
In Ref. \cite{faye04} the coefficients are given to the order $x^{11}$ inclusively.

In the Isenberg-Wilson-Mathews approach to general relativity only the conformal flat condition is employed. Thus the energy stops being analytic in $1/c$ at 3 PN. Through 2 PN order, the Isenberg-Wilson-Mathews energy of a binary is given by
\begin{eqnarray}
{\hat H}_{\rm IWM} =  -\frac{x}{2} + \bigg( \frac{3}{8} + \frac{\nu}{24} \bigg) x^2
+ \bigg( \frac{27}{16} - \frac{39}{16}\nu - \frac{17}{48} \nu^2 \bigg) x^3\,.
\end{eqnarray}
We already quote here the 3 PN result of general relativity. It reads, see Eq. (\ref{eq118}),
\begin{eqnarray}
{\hat H}_{3PN} &= & -\frac{x}{2} + \bigg( \frac{3}{8} + \frac{\nu}{24} \bigg) x^2
+ \bigg( \frac{27}{16} - \frac{19}{16}\nu + \frac{1}{48}\nu^2 \bigg) x^3 \nonumber\\[1ex]
& + &\Bigg( \frac{675}{128} + \bigg(\frac{205}{192}\pi^2-\frac{34445}{1152}\bigg)\nu
+ \frac{155}{192} \nu^2 + \frac{35}{10368} \nu^3 \Bigg) x^4 \,.
\end{eqnarray}
The difference between ${\hat H}_{\rm IWM}$ and ${\hat H}_{\rm sk}$ through 2 PN order shows the effect of truncation in the field-momentum part of ${\hat H}_{\rm sk}$  and the difference between ${\hat H}_{\rm IWM}$ and ${\hat H}_{3PN}$ reveals the effect of conformal flat truncation. In the test-body limit, $\nu =0$, all the Hamiltonians coincide and for the equal-mass case, $\nu =1/4$, their differences are largest.

\subsection{Functional representation of compact objects}	\label{sec2_3}
Before going on with the presentation of more dynamical expressions we will discuss in more detail the $\delta$-function-source model we are employing. Although we are interested in both neutron stars and black holes, our matter model will be based on black holes because these are the simplest objects in general relativity and neutron stars resemble them very much as seen from outside. The simplest black holes are the isolated non-rotating ones. Their solution is the Schwarzschild metric which solves the Einstein field equations for all time. In isotropic coordinates, the Schwarzschild metric reads
\begin{eqnarray}
ds^2 = - \left(\frac{1-{MG\over 2rc^2}}{1+{MG\over 2rc^2}}\right)^2 c^2dt^2 + \left(1+{MG\over 2rc^2}\right)^4d{\bf x}^2\,,
\end{eqnarray}
where $M$ is the gravitating mass of the black hole and $r^2 = (x^1)^2 + (x^2)^2+(x^3)^2$, $d{\bf x}^2 =  (dx^1)^2 + 
(dx^2)^2 + (dx^3)^2$. It should be pointed out that the origin of the coordinate system $r=0$ is not located where the Schwarzschild singularity $R=0$ (radial Schwarzschild coordinate R) in Schwarzschild coordinates is located, rather it is located on the other side of the Einstein-Rosen bridge, at infinity. The relation between isotropic coordinates and Schwarzschild coordinates reads $R = r (1+{MG\over 2rc^2})^2$ if $MG/2c^2 \le r$  and $R' = r (1+{MG\over 2rc^2})^2$ if $0 \le r \le MG/2c^2$, where $R'$ is another Schwarzschild radial coordinate appropriate for the geometry of the other side of the Einstein-Rosen bridge. The regimes $0 \le R < 2GM/c^2$ and $0 \le R' <  2GM/c^2$ are not accessible to isotropic coordinates. The harmonic radial coordinate, say here, $\varrho$, relates to the Schwarzschild radial coordinate through $R = \varrho + MG/c^2$. Evidently, the origin of the harmonic coordinates is located at $R = MG/c^2$ which is a spacelike curve in the region between event horizon and Schwarzschild singularity.

For two black holes, the metric for maximally sliced Brill-Lindquist initial time-symmetric data reads
\begin{eqnarray}
ds^2 = - \left(\frac{1-{\beta_1G\over 2r_1c^2}-{\beta_2G\over 2r_2c^2}}{1+{\alpha_1G\over 2r_1c^2}+{\alpha_2G\over 2r_2c^2}}\right)^2 c^2dt^2 + \left(1+{\alpha_1G\over 2r_1c^2}+{\alpha_2G\over 2r_2c^2}\right)^4d{\bf x}^2\,,
\end{eqnarray}
where the  $\alpha_a$ coefficients are given in Eq. (\ref{eq46}) and where the $\beta_a$ coefficients can be found in \cite{jaranowski02} (notice $\partial_t r_a =0$, initially).

The total energy results from the ADM mass-energy expression
\begin{eqnarray}
E_{ADM} = - \frac{c^4}{2\pi G}\oint_{i_0} ds_i \partial_i \Psi = - \frac{c^4}{2\pi G}\int d^3{\bf x} \Delta\Psi =(\alpha_1 + \alpha_2)c^2\,,\label{eq71}
\end{eqnarray}
where $\Psi = 1 + {\alpha_1G\over 2r_1c^2}+ {\alpha_2G\over 2r_2c^2}$ and $ds_i=n^ir^2d\Omega$ is a two-dimensional surface-area element with unit radial vector $n^i=x^i/r$ and solid angle element $d\Omega$.

Introducing the inversion map $r'_1 = \alpha_1^2G^2/4c^4r_1$ or, ${\bf r}'_1 =  {\bf r}_1 \alpha_1^2G^2/4c^4r_1^2$ and ${\bf r}_1 =  {\bf r}'_1 \alpha_1^2G^2/4c^4r_1'^{2}$, where 
${\bf r}_1 = {\bf x} - {\bf x}_1$, $r_1 = |{\bf x} - {\bf x}_1|$,
${\bf r}'_1 = {\bf x}' - {\bf x}_1$, $r'_1 = |{\bf x}' - {\bf x}_1|$, 
the three-metric at the throat of black hole 1, $dl^2 = \Psi^4 d{\bf x}^2$,  transforms into
\begin{eqnarray}
dl^2 = \Psi'^4 d{\bf x'}^2 = \left(1+{\alpha_1G\over
2r'_1c^2}+{\alpha_1\alpha_2G^2\over 4r_2 r'_1 c^4}\right)^4 d{\bf x'}^2\,,
\end{eqnarray}
where ${\bf r}_2 = \frac{\alpha_1^2G^2}{4c^4}\frac{{\bf r}'_1}{r_1'^{2}}
+ {\bf r}_{12}$ with ${\bf r}_{12} = {\bf r}_{1} - {\bf r}_{2}$.
From the new metric function $\Psi' = 1+{\alpha_1G\over 2r'_1c^2}+{\alpha_1\alpha_2G^2\over 4r_2r'_1c^4}$ the proper mass of the throat 1 results in, taking into account ${\bf r}_2 =  {\bf r}'_1 \alpha_1^2G^2/4c^4r_1'^{2} + {\bf r}_{12}$,
\begin{eqnarray}
m_1 &=& - \frac{c^2}{2\pi G}\oint_{i_0} ds'_i \partial'_i \Psi' = - \frac{c^2}{2\pi G}\int d^3{\bf x}' \Delta'\Psi' \\ \nonumber
&=& \alpha_1 + {\alpha_1\alpha_2G\over 2r_{12}c^2}\,.
\end{eqnarray}
This construction as performed in Ref. \cite{BL63} is a purely geometrical or vacuum one without touching singularities.
Thus having the two individual masses $m_1$ and $m_2$ the gravitational interaction energy is obtained as 
$E= E_{ADM} - (m_1+m_2)c^2$. Recall that this energy belongs to an initial value solution of the Einstein constraint equations with vanishing of both $h^{\rm TT}_{ij}$ and particle and field momenta. In this initial conditions spurious
gravitational waves are included.

Let us introduce now point masses as sources for the Schwarzschild black hole. The stress-energy tensor density of test-mass point particles in a (d+1)-dimensional curved spacetime reads
\begin{eqnarray}
{\cal{T}}_{\mu}^{\nu}(x^{\sigma}) = c^2 \sum_a m_a u_{a\mu}(t) v^{\nu}_a(t) \delta_a\,,
\end{eqnarray}
where $g^{\mu\nu}_au_{a\mu}u_{a\nu} = g_{a\mu\nu}u_a^{\mu}u_a^{\nu} = - 1$, $v_a^{\mu} = u_a^{\mu}/u_a^0$, and $\delta_a = \delta ({\bf x}-{\bf x}_a(t))$ is the usual Dirac delta functions in d-dimensional flat space. In canonical framework, the momentum density $\pi_i$ of the matter is given by
\begin{eqnarray}
\pi_i = \frac{1}{c}{\cal{T}}_{i}^{0} = c \sum_a m_a u_{ai} \delta_a\,.
\end{eqnarray}
It is the source term in the momentum constraint. The important energy density in the Hamilton constraint reads
\begin{eqnarray}
- {\cal{T}}_{\mu}^{0} n^{\mu}= - c^2 \sum_a m_a u_{a\mu} n^{\mu} \delta_a = c^2 \sum_a m_a u_a^0 N \delta_a\,,
\end{eqnarray}
where $n^{\nu}$ is the timelike unit vector, $n^{\mu}n_{\mu} = -1$ orthogonal to time slices $t$ = const, $n_{\mu} = (-N,0,0,0)$; $N$ is the lapse function, see \cite{ADM62}. Independently from the form of the metric, the equations of motion are fulfilled and can be put into the form
\begin{eqnarray}
v^{\nu}_a\nabla_{\nu}u_{a\mu} = 0\,,
\end{eqnarray}
which means geodesic motion for each particle.

The formal insertion of the stress-energy density into the Einstein field equations yields the following equations for the metric functions, for $u_{ai} = 0$,
\begin{eqnarray}
- \Psi\Delta \phi = \frac{16\pi G}{c^2} \sum_a m_a \delta_a\,,
\end{eqnarray}
where
\begin{eqnarray}
\gamma_{ij} = \Psi^{4\over d-2}\delta_{ij}, \quad  \Psi = 1 + \frac{d-2}{4(d-1)}\phi\,.
\end{eqnarray}
If the lapse function $N$ is represented by
\begin{eqnarray}
N = \frac{\chi}{\Psi}\,,
\end{eqnarray}
an equation for $\chi$ (be aware of the difference with $\chi_{\pm}$ in the Eqs. (\ref{eq64}) and (\ref{eq65})) results of the form,
\begin{eqnarray}
\Psi^2 \Delta \chi =\frac{4\pi G}{c^2} \frac{d-2}{d-1} \sum_a m_a \chi \delta_a\,.
\end{eqnarray}

With the aid of the relation,
\begin{eqnarray}
- \Delta^{-1} \delta  = \frac{\Gamma((d-2)/2)}{4\pi^{d/2}}r^{2-d}
\end{eqnarray}
it is easy to show that for $1<d<2$ the equations for $\Psi$ and $\chi$ do have well-defined solutions. Plugging in the ansatz, for d dimensions,
\begin{eqnarray}
\Psi = 1 + \frac{G(d-2)\Gamma((d-2)/2)}{c^2(d-1)\pi^{(d-2)/2}}\left({\alpha_1\over r^{d-2}_1}+ {\alpha_2\over r^{d-2}_2}\right)\,,
\end{eqnarray}
gives, for ``mass-point 1'' (mass-point seems the better notion compared with point particle or point mass because it is a fictitious particle only),
\begin{eqnarray}
\left(1 + \frac{G(d-2)\Gamma((d-2)/2)}{c^2(d-1)\pi^{(d-2)/2}}\left({\alpha_1\over
r^{d-2}_1}+ {\alpha_2\over r^{d-2}_2}\right)\right) \alpha_1 \delta_1 = m_1 \delta_1
\end{eqnarray}
or, taking $1<d<2$, and then taking the limit $r_1 \rightarrow 0$, 
\begin{eqnarray}
\left(1 + \frac{G(d-2)\Gamma((d-2)/2)}{c^2(d-1)\pi^{(d-2)/2}} {\alpha_2\over r^{d-2}_{12}}\right) \alpha_1 \delta_1 = m_1 \delta_1 \,.
\end{eqnarray}
Going over to $d=3$ by arguing that the solutions are analytic in $d$ just results in the equation, cf. Eq. (\ref{eq57}),
\begin{eqnarray}
\alpha_a = \frac{m_a}{1+ A\,\frac{\alpha_b}{r_{ab}}}\,,
\end{eqnarray}
where $b \ne a$ and $a,b = 1,2$, with the solution shown in Eq. (\ref{eq46}). The ADM energy is again given by, in the limit $d=3$, see Eq. (\ref{eq71}), 
\begin{eqnarray}
E_{ADM} =(\alpha_1 + \alpha_2)c^2\,.
\end{eqnarray}

Here we recognize the important property that although the Eqs. (\ref{eq46}) and (\ref{eq47}) may describe close binary black holes with strongly deformed apparent horizons, the both black holes can still be generated by mass-points in conformally related flat space. This is the justification for our particle model to be taken as model for orbiting black holes. We also will argue that binary black holes generated by mass-points are orbiting black holes without spin, i.e. binary Schwarzschild-type black holes. We wish to point out that at the support of our $\delta$-functions the physical spacetime is completely flat so they can not be interpreted as local sources of gravity. They rather represent wormhole geometries. The geometrical vacuum calculations in \cite{BL63} are completely finite, no infinite energies enter. The same holds with dimensional regularization where the formally infinite self-energies turn out to be zero. This is nicely consistent with the fact of the Dirac delta functions are living in flat space.

Working in the sense of distributions, the dimensional regularization procedure preserves the important law of ``tweedling of products'', \cite{infeld60}, $F_{reg}G_{reg} = (FG)_{reg}$, and gives all integrals, particularly the inverse Laplacian, a unique definition. In the sense of analytic functions, all integrals are well-defined. A famous formula derived in \cite{riesz49} plays an all-over important r{\^o}le in PN calculations,
\begin{eqnarray}
\int d^d{\bf x}~ r^{\alpha}_1r^{\beta}_2 = \pi^{d/2} \frac{\Gamma(\frac{\alpha +d}{2}) \Gamma(\frac{\beta +d}{2}) \Gamma(-\frac{\alpha+\beta +d}{2}) }{\Gamma(-\frac{\alpha}{2})\Gamma(-\frac{\beta}{2})\Gamma(\frac{\alpha +\beta+2d}{2})}r^{\alpha + \beta + d}_{12}\,.\label{eq88}
\end{eqnarray}

It is well known that distributions or generalized functions like the Dirac delta function are boundary-value functions. To overcome distributional derivations like in
\begin{eqnarray}
\partial_i \partial_j r^{2-d} = \mbox{Pf}\left((d-2)\frac{d\,n^i n^j-\delta_{ij}}{r^d}\right) - \frac{4\pi^{d/2}}{d\,\Gamma(d/2 -1)}\delta_{ij}\delta,
\end{eqnarray}
where ${\rm Pf}$ denotes the Hadamard partie finie, it is very convenient to resort on the class of analytic functions introduced in \cite{riesz49},
\begin{eqnarray}
\delta_{\epsilon} = \frac{\Gamma((d-\epsilon)/2)}{\pi^{d/2}2^{\epsilon}\Gamma(\epsilon/2)}r^{\epsilon - d} \,,
\end{eqnarray}
resulting in the Dirac delta function in the limit
\begin{eqnarray}
\delta = \mbox{lim}_{\epsilon \rightarrow 0}\delta_{\epsilon}\,.
\end{eqnarray}
On this class of functions, the inverse Laplacian operates as
\begin{eqnarray}
- \Delta^{-1} \delta_{\epsilon} =
\frac{\Gamma((d-2-\epsilon)/2)}{4\pi^{d/2}2^{\epsilon}\Gamma(\epsilon/2 + 1)} r^{\epsilon + 2  - d} = \delta_{\epsilon + 2},
\end{eqnarray}
which is special case of the convolution property $\delta_{\epsilon}*\delta_{\epsilon'} = \delta_{\epsilon+\epsilon'}$ which also results in the formula of Eq. (\ref{eq88}), and the second partial derivatives read
\begin{eqnarray}
\partial_i \partial_j r^{\epsilon + 2-d} = \mbox{Pf}\left((d-2-\epsilon)\frac{(d-\epsilon)n^i n^j-\delta_{ij}}{r^{d-\epsilon}}\right)\,.
\end{eqnarray}
No delta-function distributions are involved. Though the replacement in the stress-enery tensor density of $\delta$ through $\delta_{\epsilon}$ does destroy the divergence freeness of the stress-energy tensor and thus the integrability conditions of the Einstein theory, the relaxed Einstein field equations (the ones which result after imposing coordinate conditions) do not force the stress-energy tensor to be divergence free and can thus be solved without problems. The solutions one gets  do not fulfill the Einstein field equations but in the final limits of the $\epsilon_a$ going to zero the general coordinate covariance of the theory is recovered. This property, however, only holds if these limits were taken {\em before} the limit $d=3$ is performed, \cite{djs08}.
For completeness we give here the terms which violate the contracted Bianchi identities,
\begin{eqnarray}
\nabla_{\nu} {\cal{T}}_{\mu}^{\nu} = \frac{c^2}{2}\sum_a m_a
(g_{\rho\sigma,\mu} - (g_{\rho\sigma,\mu})_a) v^{\rho}_au^{\sigma}_a\delta_{\epsilon_a} \,.
\end{eqnarray}

We wish to point out here a difference between ADM formalism and the harmonic coordinates approach. If in the harmonic coordinates approach the stress-energy tensor is not divergence free the relaxed field equations can be solved but the harmonic coordinate conditions will not be satisfied any further. This is different with the form we use the ADM formalism where the coordinate conditions are kept valid when solving the relaxed field equations. The relaxed field equations in the harmonic case include ten functions, which are just the metric coefficients, and the divergence freeness of the stress-energy tensor is achieved if on the solution space the harmonic coordinate conditions are imposed. In the ADM formalism in Routhian form the ten metric functions do fulfill the ADM coordinate conditions and equations of motion do follow from the Routhian. They, however, will not be the ones resulting from the Einstein field equations. Those will be obtained in the limits of $\epsilon_a \rightarrow 0$ only.

The method of dimensional regularization has proven fully successful in both approaches, the Hamiltonian one and the one using the Einstein field equations in harmonic coordinates. However, another important difference between both approaches should be mentioned. Whereas in the ADM approach all poles of the type $1/(d-3)$ cancel each other and no regularization constants are left, in the harmonic gauge approach poles survive with uncancelled constants, \cite{damour01}, \cite{blanchet04a}. As found out, the difference is of gauge type only and can thus be eliminated by redefinition of the particle positions. On the other side, it shows that the positions of the mass-points in the Hamiltonian formalism are excellently chosen. Resorting to the maximally extended Schwarzschild metric, the spatial origin of the harmonic coordinates has Schwarzschild coordinate $R=MG/c^2$ inside horizon which can be reached by observers whereas the spatial origin of the ADM coordinates is located on a spacelike hypersurface at $R'=\infty$ beyond horizon. The location of the origin of the ADM coordinates allows quite a nice control of the motion of the objects.

The ADM coordinate system we are using in our article is called asymptotically maximal slicing because the trace of the extrinsic curvature of the $t$ = const spacelike slices is not zero but decays as $1/r^3$ (in four spacetime dimensions) at spacelike infinity. It is closely related with the Dirac coordinate conditions, $(\gamma^{1/3}\gamma^{ij})_{,j}=0, \, K^i_i=0$, which introduce maximal slicing. Recently, maximal slicing coordinates of the type introduced in Ref. \cite{estabrook} have proved useful in numerical relativity using moving punctures \cite{hannam07}. These coordinates are completely different from both the Dirac and ADM coordinates because those slices e.g. for Schwarschild black holes show crossings of the event horizon and settle down in the region between the Schwarzschild singularity and event horizon asymptotically. Only asymptotically these coordinates become rigidly connected to the Schwarzschild geometry. Nonetheless, moving punctures in numerical relativity are closely related with evolving Brill-Lindquist black holes.

\subsection{PN expansion of the Routh functional}	\label{sec2_4}
In case of the full Einstein theory, a formal PN expansion of the Routh functional in powers of $1/c^2$ is feasible. Using the definition $h^{\rm TT}_{ij} = \frac{16\pi G}{c^4}\hat{h}^{\rm TT}_{ij}$, we may write
\begin{eqnarray} R\left[x^i_a, p_{ai}, h^{\rm TT}_{ij}, \partial_th^{\rm TT}_{ij}\right] - c^2 \sum_a m_a  =
\sum_{n=0}^{\infty} \frac{1}{c^{2n}} R_n\left[x^i_a, p_{ai}, \hat{h}^{\rm TT}_{ij}, \partial_t{\hat{h}}^{\rm TT}_{ij}\right].
\end{eqnarray}
Furthermore, also the field equation for $h^{\rm TT}_{ij}$ can be put into a PN series form,
\begin{eqnarray}
\left(\Delta - \frac{\partial_t^2}{c^2}\right)\hat{h}^{\rm TT}_{ij} = \sum_{n=0}^{\infty} \frac{1}{c^{2n}}
D^{\rm TT}_{(n)ij}[x^k, x_a^k, p_{ak}, \hat{h}^{\rm TT}_{kl}, \partial_t{\hat{h}}^{\rm TT}_{kl}]\,.
\end{eqnarray}
This equation has to be solved iteratively with the aid of retarded integrals which themselves have to be expanded in powers of $1/c$. In higher orders, however, log-of-$1/c$ terms will show up, \cite{blanchet06}.

\subsection{Near-zone energy loss versus far-zone energy flux}	\label{sec3_5}
The change in time of the matter Hamiltonian (it is minus the Lagrangian for the gravitational field) reads, assuming ${\cal R}$ to be {\em local} in the gravitational field,
\begin{equation}
\frac{dR}{dt} = \frac{\partial R}{\partial t} =\int  \frac{\partial {\cal R}}{\partial h} {\dot h}
+ \int \frac{\partial {\cal R}}{\partial \nabla h} {\nabla \dot h}
+ \int \frac{\partial {\cal R}}{\partial {\dot h}} {\ddot h}\,,				\label{eq124}
\end{equation}
where
\begin{equation}
R = R[x^i_a, p_{ai}, h,{\dot h}] = \int {\cal R}(x^i_a, p_{ai}, h, \nabla h, {\dot h})
\end{equation}
with abbreviations
\begin{equation}
\int \equiv \int d^3{\bf x}, \quad  h \equiv h^{\rm TT}_{ij}, \quad  \nabla h
\equiv  \partial_k h^{\rm TT}_{ij},    \quad   {\dot h} \equiv  \partial_t h^{\rm TT}_{ij} \,.
\end{equation}
Above, the equation for $dR/dt$ is valid provided the equations of
motion
\begin{equation}
{\dot p}_{ai} = - \frac{\partial R}{\partial x^i_a}, \quad {\dot x}^i_a = \frac{\partial R}{\partial p_{ai}}
\end{equation}
hold. Furthermore, we have
\begin{eqnarray}
\int  \frac{\partial {\cal R}}{\partial \nabla h} {\nabla \dot h} + \int
\frac{\partial {\cal R}}{\partial {\dot h}} {\ddot h} &=& \int \nabla \left(\frac{\partial {\cal R}}{\partial \nabla h} {\dot h}\right)
 +  \frac{d}{dt} \int\left(\frac{\partial {\cal R}}{\partial {\dot h}} {\dot h}\right)\\ [2ex] \nonumber
 &-& \int  \nabla \left(\frac{\partial {\cal R}}{\partial \nabla h}\right) {\dot h}
  - \int  \frac{d}{dt}\left( \frac{\partial {\cal R}}{\partial {\dot h}}\right) {\dot h}\,.
\end{eqnarray}
Introducing the canonical field momentum
\begin{equation}
 \frac{c^3}{16\pi G} \pi =  - \frac{\partial {\cal R}}{\partial {\dot h}}\,,
\end{equation}
with abbreviation $\pi \equiv \pi^{ij}_{\rm TT}$, and the Legendre transform
\begin{equation}
H = R + \frac{c^3}{16\pi G}\int \pi {\dot h}, \quad \mbox{or} \quad R = H - \frac{c^3}{16\pi G} \int \pi {\dot h}\,,
\end{equation}
the energy loss equation takes the form
\begin{equation}
\frac{dH}{dt} = \int \nabla \left(\frac{\partial {\cal R}}{\partial \nabla h} {\dot h}\right)
 + \int \frac{\partial {\cal R}}{\partial h} {\dot h}
 - \int \nabla \left(\frac{\partial {\cal R}}{\partial \nabla h}\right) {\dot h}
  - \int \frac{d}{dt}\left( \frac{\partial {\cal R}}{\partial {\dot h}}\right) {\dot h}\,.
\end{equation}
The application of the field equations
\begin{equation}
 \frac{\partial {\cal R}}{\partial h} - \nabla \left(\frac{\partial {\cal R}}{\partial \nabla h}\right)
  - \frac{d}{dt}\left( \frac{\partial {\cal R}}{\partial {\dot h}}\right) =0
\end{equation}
results in, employing the leading order quadratic field structure of ${\cal R}$,
\begin{eqnarray}
\frac{dH}{dt} &=&  \int \nabla \left(\frac{\partial {\cal R}}{\partial
\nabla h} {\dot h}\right) = \oint_{\rm fz} d{\bf s} \frac{\partial {\cal R}}{\partial
\nabla h} {\dot h} = \frac{c^4}{32\pi G}\oint_{\rm fz} d{\bf s} (\nabla
h) {\dot h} \\[2ex] \nonumber &=& -\frac{c^3}{32\pi G}\oint_{\rm fz} d\Omega r^2 {\dot h}^2 \,,
\end{eqnarray}
where ``fz'' denotes the far zone (see e.g. section \ref{sec5_1}), $d\Omega$ the solid-angle element, and $r$ the radial coordinate of the two-surface of integration with surface-area element $d{\bf s} = {\bf n} r^2 d\Omega$. Here the further {\em assumption} has been made that the volume integrals in Eq. (\ref{eq124}) may have the outer-most region of the far zone as outer boundary. The expression
\begin{equation}
{\cal L} = \frac{c^3}{32\pi G} \oint_{\rm fz} d\Omega r^2 ({\dot h}^{\rm TT}_{ij})^2
\end{equation}
is the well known total energy flux (luminosity ${\cal L}$) of gravitational waves. The Newtonian and 1 PN wave generation fit into the above scheme of local Routhian density and far-zone as outer boundary which can be inferred from \cite{koenig03}.

\section{Binary point masses to higher PN order}	\label{sec3}
Most compact representations of dynamical systems are with Hamiltonians. Up to the 3.5 PN order, the Hamiltonian of binary point-mass systems is explicitly known, reading
\begin{eqnarray}
H(t) &=& m_1c^2 + m_2c^2 + H_N + \frac{1}{c^2} H_{[1PN]} + \frac{1}{c^4} H_{[2PN]} \nonumber\\[1ex]
&+& \frac{1}{c^5} H_{[2.5PN]}(t) + \frac{1}{c^6} H_{[3PN]} + \frac{1}{c^7} H_{[3.5PN]}(t)\,.
\end{eqnarray}
The non-autonomous dissipative Hamiltonians $H_{[2.5PN]}(t)$ and $H_{[3.5PN]}(t)$ are written as explicitly depending on time because they depend on the gravitational field variables or, in case those are reduced to matter variables, on primed matter variables, see section \ref{sec3_4}.

To simplify expressions like in section \ref{sec2_2}, we go over to the center-of-mass frame  ${\bf p}_1 + {\bf p}_2 = 0$ and also define
\begin{eqnarray}
\tilde{H} = (H-mc^2)/\mu, \quad \mu = m_1m_2/m, \quad m = m_1 + m_2, \quad \nu = \mu/m,  \nonumber \\[1ex]
{\bf p} = {\bf p}_1/\mu, \quad p_r = ({\bf n} \cdot {\bf p}), \quad {\bf q} = ({\bf x}_1 - {\bf x}_2)/Gm, \quad {\bf n} = {\bf q}/|{\bf q}|
\end{eqnarray}
with $0 \le \nu \le 1/4$ ($\nu = 0$ test-body case, $\nu = 1/4$ equal-mass case)\,.

\subsection{Conservative Hamiltonians} \label{sec3_1}
The conservative binary Hamiltonians read in reduced variables (the dissipative Hamiltonians will be treated in section \ref{sec3_4}), see \cite{djs00}, 
\begin{eqnarray}
\tilde{H}_N = \frac{p^2}{2} - \frac{1}{q}\,,
\end{eqnarray}

\begin{eqnarray}
\tilde{H}_{[1PN]} = \frac{1}{8} (3{\nu} -1) p^4 - \frac{1}{2}[(3+{\nu}) p^2 + {\nu} p^2_r]\frac{1}{q} + \frac{1}{2q^2}\,,
\end{eqnarray}
\begin{eqnarray}
\tilde{H}_{[2PN]} &=& \frac{1}{16}(1-5{\nu}+5\nu^2)p^6 \nonumber\\[1ex] &+& \frac{1}{8}[(5-20{\nu} - 3\nu^2)p^4
-2\nu^2p^2_rp^2 - 3\nu^2 p^4_r]\frac{1}{q} \nonumber\\[1ex]
&+& \frac{1}{2}[(5+8{\nu} )p^2 + 3{\nu} p_r^2]\frac{1}{q^2} - \frac{1}{4}(1+3{\nu})\frac{1}{q^3}\,,
\end{eqnarray}
\begin{eqnarray}
\tilde{H}_{[3PN]} &=& \frac{1}{128}(-5+35{\nu} - 70\nu^2 + 35 \nu^3) p^8\nonumber\\[1ex]
&+& \frac{1}{16}[(-7+42{\nu} - 53\nu^2 - 5\nu^3)p^6 +(2-3\nu)\nu^2p_r^2p^4\nonumber\\[1ex]
&+& 3(1-\nu)\nu^2p_r^4p^2 - 5\nu^3p_r^6]\frac{1}{q}\nonumber\\[1ex]
&+& [\frac{1}{16}(-27 + 136{\nu} + 109\nu^2)p^4 + \frac{1}{16}(17+30\nu){\nu} p_r^2p^2 \nonumber\\[1ex]
&+& \frac{1}{12}(5+43\nu){\nu} p_r^4]\frac{1}{q^2} \nonumber\\[1ex]
&+& [\left(-\frac{25}{8} + \left(\frac{1}{64} \pi^2 - \frac{335}{48}\right){\nu} - \frac{23}{8} \nu^2\right)p^2 \nonumber\\[1ex]
&+& \left(-\frac{85}{16} - \frac{3}{64}\pi^2 - \frac{7}{4} \nu\right){\nu} p^2_r]\frac{1}{q^3}
+[\frac{1}{8} + \left(\frac{109}{12} - \frac{21}{32} \pi^2\right) {\nu}]
\frac{1}{q^4}.
\end{eqnarray}
These Hamiltonians constitute an important element in the construction of templates for gravitational waves emitted from compact binaries. They serve also as basis of the effective one-body (EOB) approach, where with the aid of a canonical transformation the dynamics is put into test-body form of a deformed Schwarzschild metric, \cite{buonanno99}. From the reduced Hamiltonians, where a factor of $1/\nu$ is factorized out, the standard test-body dynamics is very easily obtained, simply by putting $\nu=0$.

\subsection{Dynamical invariants}	\label{sec3_2}
Dynamical invariants related to our previous dynamics are easily calculated within a Hamiltonian framework, \cite{damja00}. Let us denote the radial action by $i_r(E,j)$ with $E = {\tilde H}$ and $p^2 = p^2_r + j^2/r^2$ 
(${\bf p} = p_r {\bf e}_r +  p_{\varphi}{\bf e}_{\varphi}$ with orthonormal basis ${\bf e}_r$, ${\bf e}_{\varphi}$ in the orbital plane). Then it holds 
\begin{eqnarray}
i_r(E,j) = \frac{1}{2\pi} \oint dr ~ p_r,
\end{eqnarray}
where the integration is originally defined from minimum to minimum radial distance. Thus all
expressions derived hereof relate to orbits completed in this sense.
From analytical mechanics it is known that the phase of the completed orbit revolution $\Phi$ is given by
\begin{eqnarray}
\frac{\Phi}{2\pi} = 1 + k = - \frac{\partial}{\partial j}i_r(E,j)
\end{eqnarray}
and the orbital period $P$ reads
\begin{eqnarray}
\frac{P}{2\pi Gm} = \frac{\partial}{\partial E}i_r(E,j)\,.
\end{eqnarray}
Explicitly we get, for the periastron advance parameter $k$,
\begin{eqnarray}
k &=& \frac{1}{c^2} \frac{3}{j^2} \left\{ 1 + \frac{1}{c^2} \left[ \frac{5}{4}(7-2\nu)\frac{1}{j^2}
+ \frac{1}{2}(5-2\nu)\,E \right] \right. \nonumber \\[2ex]
&+& \left. \frac{1}{c^4} \left[a_1(\nu)\frac{1}{j^4} + a_2(\nu)\frac{E}{j^2} + a_3(\nu)\,E^2 \right]\right\}\,,
\end{eqnarray}
and for the orbital period,
\begin{eqnarray}
\frac{P}{2\pi Gm} &=& \frac{1}{(-2E)^{3/2}} \left\{1 - \frac{1}{c^2} \frac{1}{4}(15-\nu) E \right.  \nonumber \\[1ex]
&+& \left. \frac{1}{c^4} \left[\frac{3}{2}(5-2\nu)\frac{(-2E)^{3/2}}{j} -\frac{3}{32}(35+30\nu+3\nu^2)\,E^2 \right] \right. \nonumber \\[1ex]
&+& \left. \frac{1}{c^6} \left[a_2(\nu)\frac{(-2E)^{3/2}}{j^3} - 3a_3(\nu) \frac{(-2E)^{5/2}}{j} + a_4(\nu)\,E^3 \right]\right\}\,,
\end{eqnarray}
where
\begin{eqnarray}
a_1(\nu) &=& \frac{5}{2} \left(\frac{77}{2} + \left(\frac{41}{64}\pi^2-\frac{125}{3}\right)\nu + \frac{7}{4}\nu^2\right), \\[1ex]
a_2(\nu) &=& \frac{105}{2} + \left(\frac{41}{64}\pi^2-\frac{218}{3}\right)\nu + \frac{45}{6}\nu^2, \\[1ex]
a_3(\nu) &=& \frac{1}{4}(5-5\nu+4\nu^2), \\[1ex]
a_4(\nu) &=& \frac{5}{128}(21-105\nu+15\nu^2+5\nu^3).
\end{eqnarray}
These expressions have direct applications to binary pulsars, \cite{damour88}. Explicit analytical orbit solutions of the conservative dynamics through 3 PN order are given in \cite{memmesheimer}.

\subsection{ISCO and the post-Newtonian framework}	\label{sec3_3}
The motion of a test-body in the Schwarzschild metric is known to have its innermost stable circular orbit (ISCO) at $6MG/c^2$, in Schwarzschild coordinates. For test-bodies in rotating black holes (Kerr black holes) the ISCO lowers down up to $MG/c^2$ in case of direct motion and goes up to $9MG/c^2$ for retrograd motion, both motions in the equatorial plane. The ISCO of $MG/c^2$ is just the corotating case, where the Kerr black hole rotates as fast as the test-body is orbiting.
In the both limiting cases of test-body motion $MG/c^2$ and $9MG/c^2$, the black holes have maximal spins. 

Within a Hamiltonian formalism the determination of the ISCO can be done straightforwardly. Just the following two equations have to be satisfied in the center-of-mass frame within the class of orbital circles ($p_r=0)$,
\begin{eqnarray}
\frac{\partial H(r,J)}{\partial r} = 0 \quad  (\mbox{dynamical circles}), \quad \frac{\partial^2 H(r,J)}{\partial r^2} = 0\,,
\end{eqnarray}
where $r$ is the relative radial coordinate and $J$ the orbital angular momentum. With the aid of the relation for the orbital frequency
\begin{eqnarray}
\omega = \frac{dH(J)}{dJ} \,,
\end{eqnarray}
which holds for circular motion, the condition for the ISCO can also be put into the form
\begin{eqnarray}
\frac{dH(\omega)}{d\omega} = 0\,.		\label{eq114}
\end{eqnarray}
Of course, one also could work with the Legendre transform $F(\omega) = H(J) - J \omega$ and put ${d^2F(\omega)\over d\omega^2}=0$ but the outcome would be the same (recall, $dH = \omega dJ$, $dF = - J d\omega$). Stability in the present approach with circular orbits means ${d^2F\over d\omega^2} > 0$ or, ${dJ \over d\omega} < 0$ or ${dH \over d\omega} < 0$.

In the context of approximation calculations, the main difference between the $H(r,J)$ and $H(\omega)$ approaches is that $r$ is not a coordinate invariant variable; in the case of approximately known $H(r,J)$, this results in different values for the ISCO depending on the chosen coordinates. This is not the case with an approximately known $H(\omega)$ because it is coordinate invariant. Hereof, however, it does not follow that the ISCO calculated with $H(\omega)$ is more realistic than the ones calculated via $H(r,J)$ rather the varieties of ISCOs obtained via approximate $H(r,J)$ show up the uncertainty in their true location.

A test particle in the Schwarzschild spacetime on circular orbits has the reduced Hamiltonian
\begin{eqnarray}
{\hat H}(x) &=& \frac{1-2x}{(1-3x)^{1/2}} - 1 \nonumber\\[1ex]
&=& - \frac{1}{2}x + \frac{3}{8} x^2 + \frac{27}{16} x^3 + \frac{675}{128}x^4   + \frac{3969}{256}x^5  + ...\, ,
\end{eqnarray}
where
\begin{equation}
{\hat H}(x) \equiv  \frac{H (x) - mc^2}{mc^2}, \quad \quad x = \left(\frac{GM\omega}{c^3} \right)^{2/3}.
\end{equation}
The condition $\frac{d{\hat H}(x)}{dx} = 0$ yields $x=1/6 \approx 0.167$ or, in Schwarzschild coordinates $R = 6 G M/c^2$. Evidently, ISCOs are located close to the reliability limit of PN expansions, \cite{djs00}.

Using the dynamical invariants, the angular frequency of circular motion can be written as
\begin{eqnarray}
\omega_{\rm circ} = \omega_{\rm radial}+\omega_{\rm periastron} = 2\pi\frac{1+k}{P}.
\end{eqnarray}
With the aid of the definition $x=\left(\frac{GM\omega_{\rm circ}}{c^3}\right)^{2/3}$, the binary dynamics yield, through 3PN order, i.e. $c^2 {\hat H}_{3PN}  = \tilde{H}_N + \frac{1}{c^2}\tilde{H}_{[1PN]} + \frac{1}{c^4} \tilde{H}_{[2PN]} + \frac{1}{c^6}\tilde{H}_{[3PN]}$,
\begin{eqnarray}
{\hat H}_{3PN}(x)  &=& - \frac{x}{2} + \left(\frac{3}{8} + \frac{1}{24}\nu\right) x^2 + \left(\frac{27}{16} -
\frac{19}{16}\nu  +  \frac{1}{48} \nu^2\right)x^3    \nonumber\\[1ex]
&+& \left(\frac{675}{128} + \left(-\frac{34445}{1152} + \frac{205}{192}\pi^2\right) \nu +
\frac{155}{192} \nu^2 + \frac{35}{10368}\nu^3\right)x^4\,.														\label{eq118}
\end{eqnarray}
The ISCO, calculated with the aid of Eq. (\ref{eq114}), turns out to be  $x \approx 0.255$, \cite{b02}, \cite{faye04}.

To likely improve the PN truncation a representation of the energy is helpful which in the test-particle limit is known to be a ratio of two simple polynomials e.g.,
\begin{equation}
e \equiv (1 +{\hat H})^2 -1 = -x\frac{1-4x}{1-3x}
\end{equation}
For the binary 3 PN Hamiltonian, $e$ turns out to be
\begin{eqnarray}
e(x)  &=& - x[ 1- \left(1-\frac{1}{3}\nu\right)x - \left(3-\frac{35}{12}\nu\right)x^2 \\ \nonumber
&-& \left(9 +  \frac{5}{24}\left(\frac{41}{4}\pi^2 - \frac{4309}{15}\nu\right)  + \frac{103}{36} \nu^2 - \frac{1}{81}\nu^3\right)x^3 + ...].
\end{eqnarray}
Applying to this expression the technic of pad\'eing, i.e. putting it on a ratio-of-polynomials footing, the ISCO turned out to be, for equal-mass binaries, $x \approx 0.198$, \cite{djs00}.

To make contact to a recent discussion about the existence or non-existence of ISCOs for equal-mass binaries in a post-Newtonian setting \cite{blanchet03} we discuss the more general stability conditions for non-circular orbits using the Hamiltonian $H(r,p_r,J)$, where $J=p_{\varphi}$. The crucial conditions for stability are
\begin{eqnarray}
(1]: \, \frac{\partial^2 H}{\partial r^2} > 0, \quad (2]: \,
\frac{\partial^2 H}{\partial p_r^2} > 0, \quad (3]: \, \frac{\partial^2 H}{\partial r^2}\frac{\partial^2 H}{\partial p_r^2} > 0.
\end{eqnarray}
Instability occurs if one of the $>$-signs turns to zero. This particularly means that the $>$-sign in $(3]$ has to be zero. On the other side, in approximation calculations, where truncated series occur, the expression $(3]$ can be zero without one of the expressions $(1]$ or $(2]$ being zero because the product of two series of the order n PN, which is of order 2n PN, is again truncated at n PN and thus can be zero without one or both of the factors being zero. 
In Ref. \cite{blanchet03} the condition $(3]$ has been given priority because it turned out to be coordinate invariant through 3 PN order (notice in this regard, $p_r = \partial W /\partial r$, where $W$ is an action).

\subsection{PN dissipative binary dynamics}	\label{sec3_4}
The leading order 2.5 PN dissipative binary orbital dynamics is described by the non-autonomous Hamiltonian, \cite{schaefer95},
\begin{equation}
H_{[2.5PN]}(t) =
\frac{2G}{5c^5}\frac{d^3Q_{ij}(t)}{dt^3}\left(\frac{p_{1i}p_{1j}}{m_1} + \frac{p_{2i}p_{2j}}{m_2} - \frac{Gm_1m_2}{r_{12}}\right),
\end{equation}
where
\begin{equation}
Q_{ij}(t) = \sum_{a=1,2} m_a(x'^i_ax'^j_a - \frac{1}{3}{\bf x}'^2_a\delta_{ij})
\end{equation}
is the Newtonian mass-quadrupole tensor. Evidently, only after the Hamilton equations of motion are calculated the primed position and momentum variables resulting via $Q_{ij}(t)$ from time differentiations and use of the equations of motion are allowed to be identified with the unprimed position and momentum variables. The 3.5 PN Hamiltonian is known too, but it will not be given here because of quite lengthy expressions, \cite{koenig03}. Applications of the 2.5 PN Hamiltonian can be found in, e.g. \cite{kokkotas95}, \cite{ruffert96}, \cite{buonanno99}, \cite{gopu08}, where in Ref. \cite{gopu08} a transformation to the Burke-Thorne gauge (coordinate conditions) is performed.

\section{Toward binary spinning black holes}	\label{sec4}
Within the ADM formalism the action functional (i.e. the integral of the Lagrangian) of rotating bodies must have the following structure as long as the lengths of the spins are preserved in time,
\begin{eqnarray}
W &=& \int dt \Bigg( \sum_a p_{ai} \dot{x}_{a}^i + \sum_a S^{(i)}_a\Omega^{(i)}_a + \frac{1}{16 \pi} \int d^3{\bf x} \,
        \pi^{ij}_{{TT}}\dot{h}^{{TT}}_{ij} \\ \nonumber
	&-& H_{ADM}\left[{x}^i_a, p_{ai}, S^{(j)}_a, h^{{TT}}_{ij}, \pi^{ij}_{{TT}}\right]\Bigg).
\end{eqnarray}
Here, $\Omega^{(i)}_a = \Omega_{a(i)} = \frac{1}{2}\epsilon_{ijk} \Lambda_{a(l)(j)}\dot{\Lambda}_{a(l)(k)}$, $\Lambda_{a(i)(k)}\Lambda_{a(j)(k)}=\Lambda_{a(k)(i)}\Lambda_{a(k)(j)}=\delta_{ij}$, $\epsilon_{ijk}=(i-j)(j-k)(k-i)/2$; and $p_{ai}$, ${x}^i_a$, 
$S^{(i)}_a = S_{a(i)}$, and  $\delta \Theta^{(i)}_a = \frac{1}{2}\epsilon_{ijk} \Lambda_{a(l)(j)}\delta{\Lambda}_{a(l)(k)}$ are the independent matter variables, where the index $a$ again numerates the particles. Notice that $\Theta^{(i)}_a$ are anholonomic variables related to the angle-type variables $\Lambda_{a(i)(j)}$ of the proper rotations (spins). $\Omega^{(i)}_a$
is the spin precession angular frequency vector of the $a$th particle. 
The equations of motion for the particles read,
\begin{eqnarray}
\dot{x}^i_a(t) &=& \frac{\delta \int dt' H_{ADM}}{\delta p_{ai}(t)}\,, \quad
\dot{p}_{ai}(t) = - \frac{\delta \int dt' H_{ADM}}{\delta {x}^i_a(t)} \\ [1ex]
\Omega^{(i)}_a(t) &=& \frac{\delta \int dt' H_{ADM}}{\delta S^{(i)}_a(t)}\,,  \quad
{\dot{S}^{(i)}_a(t)}  = \epsilon_{ijk} \Omega^{(j)}_a(t)S^{(k)}_a(t)\,,
\end{eqnarray}
where the last equation results from the action functional through variation with respect to  $\Theta^{(i)}_a$; for more details, see e.g. \cite{hanson74}. The Hamiltonian which generates both the evolution equations for all dynamical variables as well as the contraint equations through variation with respect to the Lagrange multipliers, the lapse and shift functions $N$ and $N^i$, respectively, is given by
\begin{equation}
	H = \int d^3\mathbf{x} (N{\cal{H}} - N^i{\cal{H}}_i) + E[\gamma_{ij}],
\end{equation}
where
\begin{equation}
	E[\gamma_{ij}] = \frac{c^4}{16\pi G}\oint_{i^0} ds_i (\gamma_{ij,j} - \gamma_{jj,i})
\end{equation}
is a surface integral at spacelike infinity $i^0$ with $ds_i$ the two-dimensional surface-area element, \cite{regge74}, and
\begin{eqnarray}
	{\cal{H}}&=&{\cal{H}}^{\rm field} + {\cal{H}}^{\rm matter}  \,,\\
	{\cal{H}}_i&=&{\cal{H}}^{\rm field}_i + {\cal{H}}^{\rm matter}_i
\end{eqnarray}
with
\begin{eqnarray}
\frac{16\pi G}{c^4} {\cal{H}}^{\rm field} = - \gamma^{1/2}\mbox{R} +\frac{1}{\gamma^{1/2}}\left(\pi^i_j
\pi^j_i-\frac{1}{2}\pi^i_i\pi^j_j \right)\,,
\end{eqnarray}
\begin{eqnarray}
\frac{16\pi G}{c^3}{\cal{H}}^{\rm field}_i = 2 \partial_j \pi^j_i + \pi^{kl} \partial_i \gamma_{kl}
\end{eqnarray}
are the total Hamilton and (linear) momentum densities. After imposing the constraint equations ${\cal{H}} = {\cal{H}}_i=0$ and the coordinate conditions (\ref{eq32}) - (\ref{eq34}), the energy expression $E[\gamma_{ij}]$ turns into the ADM-Hamiltonian,
\begin{eqnarray}
E[\gamma_{ij}]=H\left[x^i_a, p_{ai}, S_a^{(i)}, h^{\rm TT}_{ij}, \pi_{\rm TT}^{ij}\right].
\end{eqnarray}

To linear order in the spin variables, the matter densities read (recall, $\delta_a=\delta({\bf x}-{\bf x}_a)$;
on simplicity reasons the index $a$ will not show up in the following equations)
\begin{eqnarray}
\mathcal{H}^{\rm matter} = - (np)c\delta - \frac{c}{2} t_{ij}^{k} \gamma^{ij}_{~~,k}
-\left[\frac{cp_l}{mc-np}\gamma^{ij}\gamma^{kl} {\hat S}_{jk}\delta\right]_{,i}\,,				\label{eq145}
\end{eqnarray}
\begin{eqnarray}
\mathcal{H}_i^{\rm matter} &=& p_i\delta + \frac{1}{2} \left[\gamma^{mk}{\hat S}_{ik} \delta\right]_{,m} \nonumber\\[1ex]
&\quad - &\left[\frac{p_lp_k}{np (mc-np)} (\gamma^{mk}\delta_i^p + \gamma^{mp} \delta_i^k)\gamma^{ql} {\hat S}_{qp}\delta\right]_{,m}\,,	\label{eq146}
\end{eqnarray}
with
\begin{eqnarray}
-np \equiv - n^{\mu}p_{\mu} = \left(m^2c^2 + \gamma^{ij}p_{i}p_{j}\right)^{1/2}
\end{eqnarray}
and
\begin{eqnarray}
t_{ij}^{k} = \gamma^{kl}\frac{\hat{S}_{l(i} p_{j)}}{np} \delta
	+ \gamma^{kl}\gamma^{mn}\frac{\hat{S}_{m(i} p_{j)} p_n p_l}{(np)^2(mc-np)} \delta\,,
\end{eqnarray}
where $n^{\mu}$ is the future directed unit vector field orthogonal to the $t$ = const hypersurfaces, 
$n_{\mu} = (- N,0,0,0)$.
The above expressions were shown to be correct up to (and including) the orders $S/c^4$ and $S/c^2$ in $\mathcal{H}^{\rm matter}$ and  $\mathcal{H}_i^{\rm matter}$, respectively.

Introducing a dreibein field $e^{(i)}_j$ with $e^{(i)}_je^{(i)}_k  = \gamma_{jk}$ and  $e_{(i)k} e_{(j)}^k  = \delta_{ij}$, a spin tensor $S_{(k)(l)} = e_{(k)}^i e_{(l)}^j {\hat S}_{ij}$ can be introduced which fulfills the relation
\begin{equation}
\gamma^{ik}\gamma^{jl}\hat{S}_{ij}\hat{S}_{kl} = 2 S_{(i)} S_{(i)} = \mbox{const},
\end{equation}
where $2S_{(i)} = \epsilon_{ijk}S_{(j)(k)}$. Crucial for our canonical formalism is the constancy in time of $S_{(i)} S_{(i)}$. Because of the symmetry property of the metric coefficients $\gamma_{ij}$, the symmetric root of $\gamma_{ij}$
\begin{equation}
	e_{il}e_{lj}= \gamma_{ij} \, , \quad e_{ij}=e_{ji}
\end{equation}
can be taken for the dreibein field, i.e. $e_{(j)k} = e_{jk}$.

To the order the formalism has been developed consistently, the following relations hold, 
\begin{equation}
p_{ai} =  \int_{V_a} d^3{\bf x} {\cal{H}}_i^{\rm matter} \,,
\end{equation}
\begin{equation}
J_{aij} = \int_{V_a} d^3{\bf x} (x^i {\cal{H}}_j^{\rm matter} - x^j {\cal{H}}_i^{\rm matter}) = x^i_a p_{aj} - x^j_ap_{ai} + S_{a(i)(j)}\,,
\end{equation}
where ${V_a}$ denotes the volume of particle $a$. Furthermore, 
\begin{equation}
\{x^i_a,p_{aj}\}=\delta_{ij}\,, \quad \{S_{a(i)}, S_{a(j)}\}= \epsilon_{ijk}S_{a(k)}\,, \quad \mbox{zero otherwise}\,,
\end{equation}
and the total linear and angular momenta respectively take the forms $P_i=\sum_a p_{ai}$ and  $J_{ij} = \sum_a J_{aij}$.

The crucial consistency relation reads,
\begin{equation}
\frac{\delta H^{\rm matter}}{\delta \gamma^{ij}} = \frac{c}{2} N \sqrt{\gamma}\,{T}_{ij}
\end{equation}
with
\begin{equation}
	H^{\rm matter} = \int d^3 \mathbf{x} (N{\cal{H}}^{\rm matter} - N^i{\cal{H}}_i^{\rm matter})\,.
\end{equation}
It is fulfilled to the needed order,
\begin{eqnarray}
\sqrt{\gamma}{T}_{ij} = - \frac{p_i p_j}{np} \delta + t_{ij,k}^k +
{\cal{O}}(G).
\end{eqnarray}

In asymptotically flat spacetimes the Poincar\'e group is a global symmetry group. Its generators $P^{\mu}$ and $J^{\mu\nu}$ are conserved and fulfill the Poincar\'e algebra, see e.g. \cite{regge74},
\begin{eqnarray}
	\{ P^{\mu} , P^{\nu} \} = 0, \\
\{ P^{\mu} , J^{\rho\sigma} \} = - \eta^{\mu\rho} P^{\sigma} + \eta^{\mu\sigma} P^{\rho}, \\
\{ J^{\mu\nu} , P^{\rho\sigma} \} = - \eta^{\nu\rho} J^{\mu\sigma} + \eta^{\mu\rho} J^{\nu\sigma}
		+ \eta^{\sigma\mu} J^{\rho\nu} - \eta^{\sigma\mu} J^{\rho\mu}.
\end{eqnarray}
The meaning of the components are energy $P^0 = H/c$, linear momentum $P^i=P_i$, angular momentum $J^{ij}=J_{ij}$, and Lorentz boost $J^{i0}/c \equiv K^i = G^i - t \, P^i$. A center-of-mass vector can be defined by $X^i = c^2 G^i / H$. 
This vector, however is not a canonical position vector, see e.g. \cite{hanson74}. The energy $H$ and the center-of-mass vector $G^i=G_i$ have the representations
\begin{eqnarray}
	H &=& - \frac{c^4}{16 \pi G} \int{{d}^3 {\bf x} \,\Delta \phi } = - \frac{c^4}{16 \pi G} \oint_{i^0}{r^2 d\Omega {\bf n}\,\nabla \phi }, \\[1ex]
	G^i &=& - \frac{c^2}{16 \pi G} \int{{d}^3{\bf x} \, x^i \Delta \phi }
= - \frac{c^2}{16 \pi G} \oint_{i^0}{r^2d\Omega n^j (x^i \partial_j - \delta_{ij})\phi },
\end{eqnarray}
where $i^{0}$ denotes spacelike infinity, $r^2 d\Omega {\bf n}$ is the two-dimensional surface-area element, and ${\bf n}$ the radial unit vector. The two quantities, $H,~G^i$, are the most involved ones of those entering the Poincar\'e algebra.

In terms of three-dimensional quantities, the Poincar\'e algebra reads, see e.g.  \cite{DJS00}, with $J_{ij} = \epsilon_{ijk}J_k$,
\begin{eqnarray}
\{ P_i , H \} = \{ J_i , H \} = 0 \, ,
\\
\{ J_i , P_j \} = \varepsilon_{ijk} \, P_k \, , \ \{ J_i , J_j \} =
\varepsilon_{ijk} \, J_k \, ,
\\
\{ J_i , G_j \} = \varepsilon_{ijk} \, G_k \, ,
\\
\{ G_i , H \} = P_i \, ,																								\label{eq165}
\\
\{ G_i , P_j \} = \frac{1}{c^2} \, H \, \delta_{ij} \, ,
\\
\{ G_i , G_j \} = - \frac{1}{c^2} \, \varepsilon_{ijk} \, J_k \, .
\end{eqnarray}
The Poincar\'e algebra has been extensively used in the calculations of PN Hamiltonians for spinning binaries, \cite{hergt08a}, \cite{hergt08b}. Hereby the most important equation was (\ref{eq165}) which tells that the total linear momentum has to be a total time derivative. Once this equation has even fixed the kinetic ambiguity in non-dimensional regularization calculations, \cite{DJS00}. The kinetic ambiguity got also fixed by a Lorentzian version of the Hadamard regularization based on the Fock-de Donder approach, \cite{bf01}.

\subsection{Approximate Hamiltonians for spinning binaries}	\label{sec4_1}
All the Hamiltonians, and the center-of-mass vectors too, given in this section have been derived or rederived in recent papers by the author and his collaborators employing general relativity in canonical form, \cite{damour08} - \cite{steinhoff08b}, adjusting it to the motion of binary black holes.

The Hamiltonian of leading-order ($LO$) spin-orbit coupling reads
\begin{equation}
	H_{{SO}}^{{LO}} = \sum_a \sum_{b \neq a} \frac{G}{c^2r_{a b}^2} (\mathbf{S}_a \times \mathbf{n}_{ab})
			\cdot \left[ \frac{3 m_b}{2 m_a} \mathbf{p}_a - 2 \mathbf{p}_b \right]
\end{equation}
and the one of leading-order spin($1$)-spin($2$) coupling is given by
\begin{equation}
	H_{{S_1S_2}}^{{LO}} = \sum_a \sum_{b \neq a} \frac{G}{2 c^2r_{a b}^3} \left[ 3 (\mathbf{S}_a \cdot \mathbf{n}_{ab})
				(\mathbf{S}_b \cdot \mathbf{n}_{ab}) - (\mathbf{S}_a \cdot \mathbf{S}_b) \right]\,,
\end{equation}
where  $ r_{ab} \mathbf{n}_{ab} =  \mathbf{x}_{a} - \mathbf{x}_{b}$, $a \ne b$ and $a,b = 1,2$. The more complicated Hamiltonian is the one with spin-squared terms because it relates to the rotational deformation of spinning black holes. To leading order, say for spin(1), it reads (details of derivation are given later)
\begin{equation}
	H_{{S^1_1}}^{{LO}} = \frac{Gm_2}{2 c^2m_1r_{12}^3} \left[ 3 (\mathbf{S}_1 \cdot \mathbf{n}_{12})
        (\mathbf{S}_1 \cdot \mathbf{n}_{12}) - (\mathbf{S}_1\cdot \mathbf{S}_1) \right] \,.										\label{eq170}
\end{equation}
The $LO$ spin-orbit and spin($a$)-spin($b$) center-of-mass vectors take the form
\begin{equation}
	\mathbf{G}_{{SO}}^{{LO}} = \sum_a \frac{1}{2c^2 m_a} (\mathbf{p}_a \times \mathbf{S}_a) \,,
	\quad \mathbf{G}_{{S_1S_2}}^{{LO}} =0 \,, \quad \mathbf{G}_{{S^2_1}}^{{LO}} = 0.
\end{equation}
Within the conservative 3 PN dynamics for spinless point masses, the center-of-mass vector has been calculated in \cite{DJS00}. Applications of the $LO$ spin Hamiltonians can be found in e.g. \cite{barker79}, \cite{d01}, \cite{schaefer04}. Other references treating the $LO$ spin dynamics are e.g. \cite{eath75}, \cite{thorne85}, \cite{kidder95}, \cite{racine08}. For applications of the next-to-leading-order ($NLO$) spin dynamics, presented straight below, see, e.g, \cite{blanchet0607}, \cite{DJS08}.

The Hamiltonian of the $NLO$ spin-orbit coupling reads, $r=r_{12}$, 
\begin{eqnarray}
H_{{SO}}^{{NLO}} &=& - G\frac{\picSin}{c^4r^2}
			\Bigg[ \frac{5 m_2 \pipi}{8 m_1^3} + \frac{3 \pipii}{4 m_1^2}
			- \frac{3 \piipii}{4 m_1 m_2}\nonumber \\ [1ex]
			&+& \frac{3 \pin \piin}{4 m_1^2}
			+ \frac{3 \piin^2 }{2 m_1 m_2} \Bigg]\nonumber\\ [1ex]
		&+& G\frac{\piicSin}{c^4r^2}
			\left[ \frac{\pipii}{m_1 m_2} + \frac{3 \pin
                        \piin }{m_1 m_2} \right]\nonumber \\ [1ex]
		&+& G \frac{\picSipii}{c^4r^2}
			\left[ \frac{2 \piin}{m_1 m_2} - \frac{3 \pin}{4 m_1^2} \right]\nonumber \\ [1ex]
		&-& G^2\frac{\picSin}{c^4r^3} \left[
			  \frac{11 m_2}{2} + \frac{5 m_2^2}{m_1}
		\right]\nonumber \\ [1ex]
		&+& G^2 \frac{\piicSin}{c^4r^3} \left[
			  6 m_1 + \frac{15 m_2}{2}
		\right]
	+ (1 \leftrightarrow 2)
\end{eqnarray}
and the one of $NLO$ spin(1)-spin(2) coupling is given by
\begin{eqnarray}
H_{{S_1S_2}}^{{NLO}} &=& \frac{G}{2 m_1 m_2 c^4r^3} [
			 6 \piicSin \picSiin\nonumber \\[1ex]
			&+& \frac{3}{2} \picSin \piicSiin \nonumber\\ [2ex]
			&-& 15 \Sin \Siin \pin \piin\nonumber \\ [2ex]
			&-& 3 \Sin \Siin \pipii
			+ 3 \Sipii \Siin \pin \nonumber\\[2ex]
			&+& 3 \Siipi \Sin \piin
			+ 3 \Sipi \Siin \piin \nonumber\\ [2ex]
			&+& 3 \Siipii \Sin \pin
			- 3 \SiSii \pin \piin \nonumber\\ [1ex]
                        &+& \Sipi \Siipii
			- \frac{1}{2} \Sipii \Siipi\nonumber
			+ \frac{1}{2} \SiSii \pipii]\nonumber \\ [1ex]
                        &+& \frac{3}{2 m_1^2 r^3} [
			- \picSin \picSiin\nonumber \\ [1ex] \nonumber
			&+& \SiSii \pin^2 - \Sin \Siipi \pin] \nonumber\\ [1ex]
		        &+& \frac{3}{2 m_2^2 r^3} [- \piicSiin \piicSin\nonumber \\ [1ex]
			&+& \SiSii \piin^2 - \Siin \Sipii \piin]
                        \nonumber \\[1ex]
		        &+& \frac{6 ( m_1 + m_2 )G^2}{c^4r^4} [ \SiSii - 2 \Sin \Siin ].
\end{eqnarray}
The calculation of the $LO$ and $NLO$ order $S_1^2$-Hamiltonians needs more information about the source terms than given in Eqs. (\ref{eq145}) and (\ref{eq146}). To achieve the $LO + NLO$ $S_1^2$-Hamiltonians, the following additional source in the Hamilton constraint is needed,
\begin{eqnarray}
\mathcal{H}^{\rm matter}_{S_1^2, {\rm static}} &=& - \frac{1}{2m_1} \left( Q^{ij}_1 \delta_1 \right)_{; ij}
+ \frac{1}{8 m_1} \gamma_{mn} \gamma^{pj} \gamma^{ql} \gamma^{mi}_{~~,p} \gamma^{nk}_{~~,q} \hat{S}_{1 ij} \hat{S}_{1 kl} \delta_1 \\ [1ex]\nonumber
	&	+& \frac{1}{4m_1} \left( \gamma^{ij} \gamma^{mn} \gamma^{kl}_{~~,m} \hat{S}_{1 ln} \hat{S}_{1 jk} \delta_1 \right)_{,i} \,,
\end{eqnarray}
where $;i$ and $,i$ denote three-dimensional covariant and partial derivatives, respectively, and where
\begin{eqnarray}
Q^{ij}_1 &\equiv& \gamma^{ik} \gamma^{jl} \gamma^{mn} \hat{S}_{1km} \hat{S}_{1nl} + \frac{2}{3} \mathbf{S}^2_1 \gamma^{ij} \,, \\ [1ex]\nonumber
2 \mathbf{S}_1^2 &=& \gamma^{ik} \gamma^{jl} \hat{S}_{1ij} \hat{S}_{1kl} = \mbox{const}.
 \end{eqnarray}
$Q^{ij}_1$ is the quadupole tensor of the black hole with number 1 resulting from its rotational deformation. Herewith, beyond the previously shown $LO$ Hamiltonian (\ref{eq170}), the NLO Hamiltonian comes out in the form, employing the Poincar\'e algebra for unique fixation of all coefficients,
 \begin{eqnarray}
H_{S_1^2}^{NLO}&=&\frac{G}{c^4r^3}\bigg[\frac{m_{2}}{4m_{1}^3}\left({\mathbf
p}_{1}\cdot{\mathbf S}_{1}\right)^2
-\frac{3}{4m_{1}m_2}{\mathbf p}_{2}^{2}{\mathbf S}_{1}^{2}
+\frac{3m_{2}}{8m_{1}^3}\left({\mathbf p}_{1}\cdot{\mathbf n}\right)^{2}{\mathbf S}_{1}^{2}
-\frac{3m_{2}}{8m_{1}^3}{\mathbf p}_{1}^{2}\left({\mathbf S}_{1}\cdot {\mathbf n}\right)^2 \nonumber \\[1ex]
&-&\frac{3m_{2}}{4m_{1}^3}\left({\mathbf p}_{1}\cdot{\mathbf n}\right)\left({\mathbf S}_{1}\cdot{\mathbf n}\right)\left({\mathbf
p}_{1}\cdot{\mathbf S}_{1}\right)
-\frac{3}{4m_{1}m_{2}}{\mathbf p}_{2}^{2}{\mathbf
S}_{1}^{2}+\frac{9}{4m_{1}m_{2}}{\mathbf p}_{2}^{2}\left({\mathbf S}_{1}\cdot{\mathbf n}\right)^2 \nonumber \\[1ex]
&+&\frac{3}{4m_{1}^2}\left({\mathbf p}_{1}\cdot{\mathbf p}_{2}\right){\mathbf S}_{1}^2
-\frac{9}{4m_{1}^2}\left({\mathbf
p}_{1}\cdot {\mathbf p}_{2}\right)\left({\mathbf S}_{1} \cdot {\mathbf n}\right)^2
-\frac{3}{2m_{1}^2}\left({\mathbf p}_{1}\cdot{\mathbf n}\right)\left({\mathbf p}_{2}\cdot{\mathbf S}_{1}\right)
\left({\mathbf S}_{1}\cdot{\mathbf n}\right)  \nonumber \\[1ex]
&+&\frac{3}{m_{1}^2}\left({\mathbf p}_{2}\cdot{\mathbf n}\right)\left({\mathbf p}_{1}
\cdot{\mathbf S}_{1}\right)\left({\mathbf S}_{1}\cdot{\mathbf n}\right)
+\frac{3}{4m_{1}^2}\left({\mathbf p}_{1}\cdot{\mathbf n}\right)\left({\mathbf p}_{2} \cdot{\mathbf n}\right){\mathbf S}_{1}^2   \nonumber \\[1ex]
&-&\frac{15}{4m_{1}^2}\left({\mathbf p}_{1}\cdot{\mathbf n}\right)\left({\mathbf p}_{2}\cdot{\mathbf n}\right)\left({\mathbf
S}_{1}\cdot{\mathbf n}\right)^2\bigg]  \nonumber \\[1ex]
&-& \frac{G^2 m_2}{2 c^4r^4} \bigg[
	9({\mathbf S}_1 \cdot {\mathbf n})^2 - 5 {\mathbf S}_1^2
	+ \frac{14 m_2}{m_1} ({\mathbf S}_1 \cdot {\mathbf n})^2  - \frac{6 m_2}{m_1}
        {\mathbf S}_1^2 \bigg]\,.
\end{eqnarray}

The spin precession equations of the Hamiltonians $H_{{SO}}^{NLO}$ and $H_{S_1^2}^{NLO}$ have been calculated also in the papers \cite{porto08a} and \cite{porto08b}, respectively, where the first paper \cite{porto08a} has benefited from paper \cite{steinhoff08a}. The final spin precession equation of the second paper \cite{porto08b}, Eq. (\ref{eq60}), deviates from the corresponding one in \cite{steinhoff08b}. A detailed inspection has shown that the last term in Eq. (\ref{eq62}) of \cite{porto08b} has wrong sign, \cite{steinhoff09}. Using the correct sign, after redefinition of the spin variable, agreement with the Hamiltonian of Ref. \cite{steinhoff08b} is achieved.

The $NLO$ order spin-orbit and spin($a$)-spin($b$) center-of-mass vectors take the form
\begin{eqnarray}
\mathbf{G}_{{SO}}^{{NLO}} &=& - \sum_a \frac{\mathbf{p}_a^2}{8 c^4m_a^3} (\mathbf{p}_a \times \mathbf{S}_a)  \nonumber \\[1ex]
		&+& \sum_a \sum_{b \neq a} \frac{m_bG}{4 c^4m_a r_{ab}} \bigg[ ((\mathbf{p}_a \times \mathbf{S}_a) \cdot \mathbf{n}_{ab})
			\frac{5\mathbf{x}_a+\mathbf{x}_b}{r_{ab}} - 5 (\mathbf{p}_a \times \mathbf{S}_a) \bigg]  \nonumber \\[1ex]
	&+& \sum_a \sum_{b \neq a} \frac{G}{c^4 r_{ab}} \bigg[ \frac{3}{2} (\mathbf{p}_b \times \mathbf{S}_a)
			- \frac{1}{2} (\mathbf{n}_{ab} \times \mathbf{S}_a) (\mathbf{p}_b \cdot \mathbf{n}_{ab}) \nonumber\\[1ex]
			&-& ((\mathbf{p}_a \times \mathbf{S}_a) \cdot \mathbf{n}_{ab}) \frac{\mathbf{x}_a+\mathbf{x}_b}{r_{ab}} \bigg],
\end{eqnarray}
\begin{equation}
\mathbf{G}_{{S_1S_2}}^{NLO} = \frac{G}{2c^4} \sum_a \sum_{b \neq a} \bigg\{
	\left[3(\mathbf{S}_{a}\cdot\mathbf{n}_{ab})(\mathbf{S}_{b}\cdot\mathbf{n}_{ab}) -(\mathbf{S}_{a}\cdot\mathbf{S}_{b})\right]
		\frac{\mathbf{x}_{a}}{r_{ab}^3} + (\mathbf{S}_{b}\cdot\mathbf{n}_{ab}) \frac{\mathbf{S}_{a}}{r_{ab}^2} \bigg\}\,,
\end{equation}
\begin{equation}
\mathbf{G}_{S_{1}^2}^{NLO}
 =\frac{2m_{2}G}{c^4m_{1}}\left[\frac{3\left(\mathbf{S}_{1}\cdot\mathbf{n}_{12}\right)^2}{8r_{12}^3}\left(\mathbf{x}_{1}+\mathbf{x}_{2}\right)
 +\frac{\mathbf{S}_{1}^2}{8r_{12}^3} \left(3\mathbf{x}_{1}-5\mathbf{x}_{2}\right)
-\frac{\left(\mathbf{S}_{1}\cdot\mathbf{n}_{12}\right)\mathbf{S}_1}{r_{12}^2}\right]\,,
\end{equation}
summing up through 2 PN order in the spin parts, we obtain, also see \cite{DJS00},
\begin{equation}
\mathbf{G}=\mathbf{G}_{N}+\mathbf{G}_{1PN}+\mathbf{G}_{2PN}+\mathbf{G}_{3PN}+       \mathbf{G}_{SO}^{LO}+\mathbf{G}_{SO}^{NLO}
 +\mathbf{G}_{S_{1}S_{2}}^{NLO} +\mathbf{G}_{S_{1}^2}^{NLO} + \mathbf{G}_{S_{2}^2}^{NLO}\,.
 \end{equation}
Numerically, spins of black holes can be counted of order $1/c$ (maximum), thus the spinless parts
are taken up to the 3 PN order.

The currently known conservative binary Hamiltonians for spinning black holes through order $1/c^4$ can be summarized as follows
\begin{eqnarray}
H &=& H_{N}+H_{1PN}+H_{2PN}+H_{3PN}  \nonumber \\[1ex]
&+& H_{SO}^{LO}+H^{LO}_{S_{1}S_{2}} + H^{LO}_{S_{1}^2}+ H^{LO}_{S_{2}^2} \nonumber \\[1ex]
&+& H_{SO}^{NLO}+H^{NLO}_{S_{1}S_{2}} + H^{NLO}_{S_{1}^2}+ H^{NLO}_{S_{2}^2} \nonumber \\ [1ex]
 &+& H_{p_{1}S_{2}^3}+H_{p_{2}S_{1}^3} + H_{p_{1}S_{1}^3}+H_{p_{2}S_{2}^3} \nonumber \\[1ex]
&+& H_{p_{1}S_{1}S_{2}^2} + H_{p_{2}S_{2}S_{1}^2} + H_{p_{1}S_{2}S_{1}^2} + H_{p_{2}S_{1}S_{2}^2} \nonumber \\[1ex]
&+&  H_{S_{1}^2S_{2}^2} + H_{S_{1} S_{2}^3} + H_{S_{2}S_{1}^3}\,.
\end{eqnarray}
The Hamiltonians $H_{S_{1}^4}$ and $H_{S_{2}^4}$ in the approximation in question turned out to be zero.

\section{Lorentz-covariant approach and PN expansions}	\label{sec5}
The Lorentz-covariant approach has found a quite thorough presentation in \cite{blanchet06}. So we will not go into so many details as in the canonical approach presented in the previous sections.

The Einstein field equations are given by
\begin{equation}
G^{\mu\nu}(g_{\kappa\lambda}, \partial_{\alpha}g_{\kappa\lambda}, \partial_{\alpha}\partial_{\beta}g_{\kappa\lambda})
= \frac{8\pi {G}}{c^2} \frac{T^{\mu\nu}(g_{\kappa\lambda}; c^2)}{{c^2}}\,,
\end{equation}
where $g_{\kappa\lambda}$ and $T^{\mu\nu}$  are the four-metric and the stress-energy tensor of the matter (e.g., fluid), respectively. The contracted Bianchi identities yield the four-dimensional equations of motion (EOM) for the matter,
\begin{equation}
\nabla_{\nu}G^{\mu\nu} \equiv 0 \quad \rightarrow \quad \nabla_{\nu}T^{\mu\nu} = 0 \quad \mbox {(EOM)},
\end{equation}
where $\nabla_{\nu}$ denotes the four-dimensional covariant derivative. The Landau-Lifshitz form of the Einstein field equations fits very well into Lorentz-covariant schemes. It takes the form, \cite{landau85},
\begin{equation}
\partial_{\lambda}\partial_{\kappa} U^{\mu\nu\lambda\kappa}(g_{\alpha\beta})
= \frac{16\pi G}{c^4} \tau^{\mu\nu}_{\rm LL}(g_{\alpha\beta}, \partial_{\gamma}g_{\alpha\beta})\,,
\end{equation}
with $U^{\mu\nu\lambda\kappa} = {\bf g}^{\mu\nu} {\bf g}^{\lambda\kappa} - {\bf g}^{\mu\lambda}{\bf g}^{\nu\kappa}$ and  ${\bf g}^{\mu\nu} = \sqrt{-g} g^{\mu\nu}$, where $g$ denotes the determinant of the metric tensor. $ \tau^{\mu\nu}_{\rm LL}$ is known as the Landau-Lifshitz stress-energy (or energy-momentum) pseudo-tensor of the gravitational field. It is unique in the sense of symmetry and dependence on the metric coefficients and its first derivatives only. The equations of motion now read
\begin{equation}
\partial_{\nu}\partial_{\lambda}\partial_{\kappa}U^{\mu\nu\lambda\kappa}
\equiv 0 \quad \rightarrow \quad \partial_{\nu} \tau^{\mu\nu}_{\rm LL} = 0 \quad \mbox {(EOM)}
\end{equation}
with
\begin{equation}
\tau^{\mu\nu}_{\rm LL} \equiv -g  T^{\mu\nu} + \frac{c^4}{16\pi G} t^{\mu\nu}_{\rm LL}(g_{\alpha\beta}, \partial_{\gamma} g_{\alpha\beta})\,.
\end{equation}

Applying the condition of harmonic coordinates
\begin{equation}
\partial_{\nu}{\bf g}^{\mu\nu} = 0 \quad \mbox{or}, \quad \partial_{\nu}H^{\mu\nu}
= 0  \quad \mbox{with} \quad H^{\mu\nu} = \sqrt{-g}g^{\mu\nu} - \eta^{\mu\nu}\,,
\end{equation}
where $ \eta^{\mu\nu}$ denotes the Minkowski metric, the Einstein field equations in Landau-Lifshitz form read (so-called relaxed field equations because they do not imply the equations of motion; rather the condition of harmonic coordinates implies the equations of motion),
\begin{equation}
\eta^{\alpha\beta}\partial_{\alpha}\partial_{\beta}H^{\mu\nu} = \frac{16\pi G}{c^4} \tau^{\mu\nu}, \label{eq188}
\end{equation}
where
\begin{equation}
\tau^{\mu\nu} = - g  T^{\mu\nu} + \frac{c^4}{16\pi G} \Lambda^{\mu\nu}									
\end{equation}
and
\begin{eqnarray}
\Lambda^{\mu\nu} &=& -  H^{\alpha\beta}\partial_{\alpha}\partial_{\beta} H^{\mu\nu} + \partial_{\alpha}H^{\mu\beta}\partial_{\beta}H^{\nu\alpha} + \frac{1}{2} g^{\mu\nu}g_{\alpha\beta} \partial_{\lambda} H^{\alpha\tau}
\partial_{\tau} H^{\beta\lambda} \\  \nonumber
&+& \frac{1}{8}(2g^{\mu\alpha}g^{\nu\beta} - g^{\mu\nu}g^{\alpha\beta})(2g_{\lambda\tau}g_{\rho\sigma} -
g_{\lambda\sigma}g_{\tau\rho})\partial_{\alpha} H^{\lambda\sigma} \partial_{\beta} H^{\tau\rho} \\ [1ex]\nonumber
&-& g^{\mu\alpha}g_{\beta\tau} \partial_{\lambda}H^{\nu\tau} \partial_{\alpha}H^{\beta\lambda}
- g^{\nu\alpha} g_{\beta\tau} \partial_{\lambda}H^{\mu\tau} \partial_{\alpha}H^{\beta\lambda}
+ g_{\alpha\beta} g^{\lambda\tau} \partial_{\lambda} H^{\mu\alpha} \partial_{\tau}H^{\beta\nu}.
\end{eqnarray}
The $\Lambda^{\mu\nu}$ object starts with quadratic non-linearities of the gravitational field. It is another stress-energy pseudo-tensor of the gravitational field.

\subsection{PM and PN expansions}	\label{sec5_1}
The {\em formal} retarded solution (resulting under the condition of no incoming radiation) of the inhomogeneous wave equation (\ref{eq188}) reads
\begin{equation}
H^{\mu\nu}({\bf x},t) = - \frac{4G}{c^4} \int d^3{\bf x}'\tau^{\mu\nu}({\bf x'},t - \frac{|{\bf x}- {\bf x'}|}{c};c^2) |{\bf x}- {\bf x'}|^{-1}\,.
\end{equation}
A post-Minkowskian (PM) expansion in powers of $G$ can now be introduced in the form,
\begin{equation}
H^{\mu\nu}({\bf x},t) =
\sum_{n=1}^{\infty} G^n H^{\mu\nu}_{[n]}({\bf x},t;c)\,.
\end{equation}
If additionally the virial theorem holds,
\begin{eqnarray}
\frac{{G}M}{Rc^2} \sim \frac{V^2}{{c^2}} \,,
\end{eqnarray}
where respectively $M$, $R$, and $V$ are typical masses, radii, and velocities of the system in question, the PM expansion may be further expanded into a PN series in powers of $1/c$. Yet, because of the retardation structure of the solution, a PN expansion is achieveable only in the near and far zones and this even only in a generalized form with log-$c$ terms showing up at higher orders starting from 4 PN, i.e. $(1/c^2)^4=1/c^8$, on.

Let us assume now that the matter source is bounded by a sphere with radius $R$ centered in the origin of the coordinate system and that for the typical gravitational wave length $\lambda$ the relation  $\lambda >> R$ holds. The near zone is then defined by $|{\bf x}| << \lambda$. The {\em formal} PN expansion (near-zone PN expansion) is defined by
\begin{equation}
H^{\mu\nu}_{\rm nz}({\bf x},t) = - \frac{4G}{{c^4}} \sum_{n=0}^{\infty}
\frac{(-1)^n}{n!}\int d^3{\bf x}' \frac{\partial^n}{{c^n}\partial t^n} \tau^{\mu\nu}({\bf x'},t;c^2) |{\bf x}- {\bf x'}|^{n-1}\,. 	\label{eq194}
\end{equation}
Additionally, the expansion
\begin{equation}
\tau^{\mu\nu}({\bf x'},t;c^2) = \sum_{n=-1}^{\infty} \frac{1}{c^{2n}}\tau_{(n)}^{\mu\nu}({\bf x'},t)							\label{eq195}
\end{equation}
applies.
In the far zone, where  $r = |{\bf x}| >> \lambda$ holds, a {\em formal} PN expansion (far-zone PN expansion) yields
\begin{equation}
H^{\mu\nu}_{\rm fz}({\bf x},t) = - \frac{4G}{{c^4}r} \sum_{n=0}^{\infty} \frac{1}{n!} \int d^3{\bf x}' \frac{\partial^n}{{c^n}\partial t^n}
\tau^{\mu\nu}({\bf x'},t-\frac{r}{{c}};c^2) ({\bf x'} \cdot {\bf n})^n\,.														\label{eq196}
\end{equation}
Here, the expansion
\begin{equation}
\tau^{\mu\nu}({\bf x'},t-\frac{r}{c};c^2) = \sum_{n=-1}^{\infty} \frac{1}{c^{2n}}\tau_{[n]}^{\mu\nu}({\bf x'},t -\frac{r}{c})	\label{eq197}
\end{equation}
applies. The Eqs. (\ref{eq194}) and (\ref{eq195}) on the one side and (\ref{eq196}) and (\ref{eq197}) on the other are somewhat simplified in the sense that they do not show up log-$c$ terms at higher orders in the expansions which result from the badly defined integrals of non-compact support. On the other side, as the expressions stand, they are mathematically not defined at all. Details can be found in the works by Blanchet, Damour, Will, and collaborators; particularly see the contribution by L. Blanchet in this volume; for tail terms, also see our section \ref{sec5_3}.

\subsection{PN expansion in the near zone}	\label{sec5_2}
Up to the 2 PN order the metric coefficients read
\begin{eqnarray}
g_{00} &=& -1+\frac{2}{c^2} V-\frac{2}{c^4} V^2+ \frac{8}{c^6} \left( \hat{X}+V_i V_i+\frac{V^3}{6} \right),  \\[1ex]
g_{0i} &=& -\frac{4}{c^3} V_i -\frac{8}{c^5} \hat{R}_i, \\[1ex]
g_{ij} &=& \delta_{ij} \left[ 1+\frac{2}{c^2} V+\frac{2}{c^4} V^2 \right]+
\frac{4}{c^4} \hat{W}_{ij}.
\end{eqnarray}
With the following choice of the matter variables, respectively mass, mass-current, and stress density,
\begin{equation}
\sigma = \frac{T^{00}+T^{ii}}{c^2}\,,\quad \sigma_i = \frac{T^{0i}}{c}\,, \quad \sigma_{ij} = T^{ij}\,,
\end{equation}
the 2 PN potentials can be put into the form
\begin{equation}
V({\bf x},t) = G_{\rm ret}\left\{-4\pi G \sigma \right\} \equiv G \int {d^3{\bf z}\over |{\bf x}-{\bf z}|}
\sigma( {\bf z}, t - |{\bf x} -{\bf z}|/c ) \,,
\end{equation}

\begin{equation}
V_i = G_{\rm ret}\left\{-4\pi G \sigma_i\right\}\,,
\end{equation}

\begin{equation}
{\hat W}_{ij} =  G_{\rm ret}\left\{-4 \pi G(\sigma_{ij} - \delta_{ij} \sigma_{kk}) - \partial_i V \partial_j V\right\}\,,
\end{equation}

\begin{equation}
{\hat R}_i =   G_{\rm ret}\left\{ - 4\pi G(V\sigma_i - V_i \sigma) - 2 \partial_k V
\partial_i V_k - {3\over 2} \partial_t V \partial_i V \right\}\,,
\end{equation}

\begin{eqnarray}
{\hat X} &=&   G_{\rm ret}\biggl\{ -4\pi GV \sigma_{ii} + 2 V_i \partial_t \partial_i V +V \partial_t^2 V  \nonumber \\
&+& {3\over 2} (\partial_t V)^2 - 2 \partial_i V_j \partial_j V_i + \hat{W}_{ij} \partial^2_{ij} V \biggr\}\,.
\end{eqnarray}

The potentials of the orders 2.5 PN and 3.5 PN are radiation-reaction potentials. They are most compactly given under Burke-Thorne coordinate conditions, reading
\begin{eqnarray}
U^{\rm reac}(\mathbf{x},t) &=& - \frac{G}{5c^5} x^{ij} {\hat {\rm M}}^{[5]}_{ij}(t) +
\frac{G}{c^7} \bigg[ \frac{1}{189} x^{ijk} {\hat {\rm M}}^{[7]}_{ijk}(t)  \nonumber
\\ &~& - \frac{1}{70} x^{kk} x^{ij} {\hat {\rm M}}^{[7]}_{ij}(t) \bigg]\,, \\[1ex]
U^{\rm reac}_i(\mathbf{x},t) &=&~ \frac{G}{c^5} \bigg[\frac{1}{21} \hat{x}^{ijk} {\hat {\rm M}}^{[6]}_{jk}(t) - \frac{4}{45} \epsilon_{ijk} x^{jm}
{\hat {\rm S}}^{[5]}_{km}(t) \bigg]\,,
\end{eqnarray}
where the source multipole moments are given by
\begin{equation}
{\hat {\rm M}}_{ij} = \int d^3\mathbf{y} \left(\hat{y}^{ij} \sigma+ \frac{1}{14c^2}
  y^{kk} \hat{y}^{ij} \partial_t^2 \sigma - \frac{20}{21c^2} \hat{y}^{ijk} \partial_t \sigma_k \right)\,,
\end{equation}
\begin{equation}
{\hat {\rm M}}_{ijk} = \int d^3\mathbf{y}~\hat{y}^{ijk} \sigma\,,
\end{equation}
\begin{equation}
{\hat {\rm S}}_{ij} = \int d^3\mathbf{y}~ \epsilon_{km<i} \hat{y}^{j>k} \sigma_m\,.
\end{equation}
The used definitions read $y^{ij} \equiv y^i y^i$, $y^{<ij>} \equiv \hat{y}^{ij} = {\rm STF} (y^{ij})$, and e.g.,  ${\hat {\rm M}}^{[7]}_{ijk}$ indicates the seventh time derivative of ${\rm M}_{ijk}$. Explicitly, the 1 PN metric including the gravitational radiation reaction through 3.5 PN order is given by
\begin{eqnarray}
\mbox{} \! \! \! g_{00} &=& -1 + \frac{2}{c^2} (U+U^{\rm reac}) +\frac{1}{c^4} \left[\partial_t^2 \chi -2 U^2 - 4 U U^{\rm reac} \right], \\
\mbox{} \! \! \! g_{0i} &=& -\frac{4}{c^3} (U_i+U_i^{\rm reac}), \\
\mbox{} \! \! \! g_{ij} &=& ~ \delta_{ij} \left[1+ \frac{2}{c^2} (U+U^{\rm reac})\right] ,
\end{eqnarray}
where the potentials have the integral representations
\begin{eqnarray}
 U({\mathbf{x},t}) &=& G  \int \frac{d^3\mathbf{y}}{|\mathbf{x}-\mathbf{y}|}~\sigma({{\bf y},t}), \\  U_i({\mathbf{x},t}) &=& G \int
 \frac{d^3\mathbf{y}}{|\mathbf{x}-\mathbf{y}|}~ \sigma_i({\mathbf{y},t}),\\
\chi({\mathbf{x},t}) &=& G \int d^3\mathbf{y} ~|\mathbf{x}-\mathbf{y}| \sigma({\bf y},t).
\end{eqnarray}
These integrals are well defined. Evidently, multipole expansion and PN expansion nicely fit together;
see also \cite{poujade02}.

\subsection{PN expansion in the far zone}	\label{sec5_3}
In the far zone, the multipole expansion of the transverse-traceless (TT) part of the gravitational field, 
obtained by algebraic projection with $P_{ijkm}(\mathbf{n})$, reads, e.g. \cite{thorne80}, 
\begin{eqnarray}
H^{ij}_{\rm fzTT}(\mathbf{x},t) &=& - \frac{G}{{c^4}}\frac{P_{ijkm}
(\mathbf{n})}{r} \sum_{l=2}^{\infty} \left\{\left(\frac{1}{{c^2}}\right)^{\frac{l-2}{2}} \frac{4}{l!}~
\mbox{M}^{[l]}_{kmi_3...i_l}(t-\frac{r_*}{{c}})~N_{i_3...i_l} \right. \nonumber \\[2ex]
&+& \left. \left(\frac{1}{{c^2}}\right)^{\frac{l-1}{2}} \frac{8l}{(l+1)!}~
\epsilon_{pq(k}~\mbox{S}^{[l]}_{m)pi_3...i_l}(t-\frac{r_*}{{c}})~n_q~N_{i_3...i_l}\right\},							\label{eq218}
\end{eqnarray}
where the leading mass-quadrupole tensor takes the form, e.g. \cite{BS93}, 
\begin{eqnarray}
\mbox{M}_{ij}(t-\frac{r_*}{c}) &=& \widehat{\mbox{M}}_{ij}\left(t - \frac{r_*}{c}\right) \nonumber \\[2ex] +~
\frac{2Gm}{c^3} \int_{0}^{\infty} &dv& \left[\mbox{ln} \left(\frac{v}{2b}\right) + \frac{11}{12}\right]
\widehat{\mbox{M}}^{[2]}_{ij}(t-\frac{r_*}{c}-v) + {\cal O}\left(\frac{1}{c^4}\right) \label{eq219}
\end{eqnarray}
with
\begin{eqnarray}\nonumber
r_* = r + \frac{2Gm}{c^2}\mbox{ln}\left(\frac{r}{cb}\right) + {\cal O}\left(\frac{1}{c^3}\right)
\end{eqnarray}
showing a leading-order tail term. Notice the modification of the standard PN expansion through tail terms. The Eq. (\ref{eq218}) nicely shows that also multipole expansions in the far zone do induce PN expansions.

The gravitational luminosity is generally given by ($H_{\rm fzTT}^{ij} = - h^{\rm TTfz}_{ij}$),
\begin{eqnarray}
{\cal{L}}(t) = \frac{c^3}{32\pi G} \oint_{\rm fz} (\partial_tH_{\rm fzTT}^{ij})^2 r^2d\Omega \,.
\end{eqnarray}
Through 1.5 PN order, the luminosity explicitly reads,
\begin{eqnarray}
{\cal{L}}(t) &=& \frac{G}{5c^5}\sum_{n=0}^{\infty}\left(\frac{1}{c^2}\right)^n {\hat{\cal{L}}}_{n}(t) \nonumber \\
&=& \frac{G}{5c^5}\left\{\mbox{M}^{[3]}_{ij}\mbox{M}^{[3]}_{ij} +
\frac{1}{c^2}\left[\frac{5}{189}\mbox{M}^{[4]}_{ijk}\mbox{M}^{[4]}_{ijk}
+ \frac{16}{9}\mbox{S}^{[3]}_{ij}\mbox{S}^{[3]}_{ij}\right] \right\}.
\end{eqnarray}
On reasons of energy balance, for any representation of the Einstein theory, the time-averaged energy loss has to fulfill a relation of the form 
\begin{eqnarray}
-<\frac{d{\cal{E}}(t-\frac{r_*}{c})}{dt}> ~=~ <{\cal{L}}(t)>\,,													\label{eq222}
\end{eqnarray}
where the time averaging procedure takes into account typical periods of the system. The derivation of this equation in 
section \ref{sec3_5} is known to be valid for the first two radiation emission and reaction levels.

\section{Energy loss and gravitational wave emission}	\label{sec6}
The energy flux to n PN order in the far zone, denoted n PN(fz), implies energy loss to (n+5/2) PN order
in the near zone, denoted (n+5/2) PN(nz). Hereof it follows that energy-loss calculations are quite efficient via energy-flux calculations. Because of this we will apply the balance property between emitted and lost energies to some PN orders to easily derive the energy loss from the energy flux. In general, only after averaging over orbital periods the both expressions will coincide (see Eq. (\ref{eq222})).  In the case of circular orbits, however, this averaging procedure is not needed.

\subsection{Orbital decay to 4 PN order}	\label{sec6_1}
The binding energy of our binary system on circular orbits is given by $\mu E_{\rm circ}$. Therefore, for the energy loss to 4 PN order, we get
\begin{eqnarray}
- \mu \frac{dE_{\rm circ}}{dt} = {\cal{L}} &=& \frac{32c^5}{5G} \nu^2x^5 \left[1-\left(\frac{1247}{336} 
+ \frac{35}{12}\nu\right)x + 4\pi x^{3/2}\right],
\end{eqnarray}
where the 1.5 PN(fz) energy flux is taken from Ref. \cite{BDI95} where also the 2 PN(fz) energy flux can be found; 
for the 3.5 PN(fz) energy flux see \cite{BFIJ02}, \cite{BIJ02}.

Taking into account the Eq. (\ref{eq118}) we obtain a differential equation for $x$ which is easily solved with accuracy $1/c^8$. In terms of the dimensionless time variable
\begin{eqnarray}
\tau= \frac{\nu c^3}{5Gm}(t_c-t),
\end{eqnarray}
where $t_c$ denotes the coalescence time, the solution reads \cite{BDI95},
\begin{eqnarray}
x &=& \frac{1}{4} \tau^{-1/4} \left[1+\left(\frac{743}{4032}
+ \frac{11}{48}\nu\right)\tau^{-1/4} - \frac{1}{5}\pi \tau^{-3/8} \right].
\end{eqnarray}

Taking into account the relation between phase and frequency
$\frac{d\phi}{dt} = \omega$, respectively  $\frac{d\phi}{d\tau} = -\frac{5}{\nu}x^{3/2}$,
the phase evolution results in
\begin{eqnarray}
\phi &=& \phi_c  - \frac{1}{\nu} \tau^{5/8}\left[1+\left(\frac{3715}{8064}
+ \frac{55}{96}\nu\right)\tau^{-1/4} - \frac{3}{4}\pi \tau^{-3/8} \right].
\end{eqnarray}

\subsection{Gravitational waveform to 1.5 PN order}	\label{sec6_2}
The radiation field can be decomposed into two orthogonal polarization states. The polarization states $h_+$ and $h_{\times}$ are defined by
\begin{eqnarray}
h_+ &=& \frac{1}{2}(u_iu_j-v_iv_j)h^{\rm TT}_{ij}, \\
h_{\times} &=& \frac{1}{2}(u_iv_j+v_iu_j)h^{\rm TT}_{ij},
\end{eqnarray}
where $\mathbf{u}$ and $\mathbf{v}$ denote two vectors in the polarization plane forming an orthogonal right-handed triad with the direction $\mathbf{n}$ from the source to the detector. The detector is directly sensitive to a linear combination of the polarization waveforms $h_+$ and $h_{\times}$, namely
\begin{eqnarray}
h(t) = F_+h_+(t) + F_{\times} h_{\times}(t),
\end{eqnarray}
where $F_+$ and  $F_{\times}$ are the so-called beam-pattern functions of the detector depending on two angles giving the direction $-\mathbf{n}$ of the source as seen from the detector and a polarization angle specifying the orientation of the vectors $\mathbf{u}$ and $\mathbf{v}$ around that direction.

For our binary system, the two polarizations  $h_+$ and  $h_{\times}$ are chosen such that the polarization vectors  $\mathbf{u}$ and $\mathbf{v}$ lie respectively along the major and minor axis of the projection onto the plane of the sky of the circular orbit, with  $\mathbf{u}$ oriented toward the ascending node, the point at which black hole 1 crosses the plane of the sky moving toward the observer. The result, to 1.5 PN(fz) order, reads \cite{BIWW96} (the 2 PN(fz) wave form is given therein too)
\begin{eqnarray}
h_{+,\times} = \frac{2G\mu x}{c^2r}\left[H^{[0]}_{+,\times} + x^{1/2}H^{[1/2]}_{+,\times}
+ xH^{[1]}_{+,\times} + x^{3/2}H^{[3/2]}_{+,\times}\right],
\end{eqnarray}
where the plus polarization is given by
\begin{eqnarray}
H^{[0]}_{+} &=& - (1+c_i^2)\mbox{cos}2\psi, \\[1ex]
H^{[1/2]}_{+} &=& - \frac{s_i}{8} \frac{\delta m}{m}[(5+c_i^2)\mbox{cos}\psi - 9(1+c_i^2)\mbox{cos}3\psi], \\[1ex]
H^{[1]}_{+} &=& \frac{1}{6}[19+19c_i^2-2c_i^4-\nu(19-11c^2_i-6c_i^4)]\mbox{cos}2\psi \nonumber \\[1ex]
&-& \frac{4}{3}s_i^2(1+c_i^2)(1-3\nu)\mbox{cos}4\psi\,, \\[1ex]
H^{[3/2]}_{+} &=& \frac{s_i}{192}\frac{\delta m}{m} \{[57+60c_i^2-c_i^4-2\nu(49-12c_i^2-c_i^4)] \mbox{cos}\psi \nonumber \\[1ex]
&-& \frac{27}{2}[73+40c_i^2-9c_i^4-2\nu(25-8c_i^2-9c_i^4)]\mbox{cos}3\psi \nonumber \\[1ex]
&+& \frac{625}{2}(1-2\nu)s_i^2(1+c_i^2)\mbox{cos}5\psi\} -2\pi(1+c_i^2)\mbox{cos}2\psi\,,
\end{eqnarray}
and the cross polarization by
\begin{eqnarray}
H^{[0]}_{\times} &=& - 2c_i\mbox{sin}2\psi\,,\\[1ex]
H^{[1/2]}_{\times} &=&- \frac{3}{4} s_ic_i\frac{\delta m}{m}[\mbox{sin}\psi - 3\mbox{sin}3\psi]\,, \\[1ex]
H^{[1]}_{\times} &=& \frac{c_i}{3}[17-4c_i^2-\nu(13-12c^2_i)]\mbox{sin}2\psi - \frac{8}{3}c_is_i^2(1-3\nu)\mbox{sin}4\psi\,, \\[1ex]
H^{[3/2]}_{\times} &=& \frac{s_ic_i}{96}\frac{\delta m}{m} \{[63-5c_i^2-2\nu(23-c_i^2)] \mbox{sin}\psi  \nonumber \\[1ex]
&-& \frac{27}{2}[67-15c_i^2-2\nu(19-15c_i^2)]\mbox{sin}3\psi  \nonumber \\[1ex]
&+& \frac{625}{2}(1-2\nu)s_i^2\mbox{sin}5\psi\} -4\pi c_i \mbox{sin}2\psi\,,
\end{eqnarray}
where $c_i=\mbox{cos}i$ and $s_i=\mbox{sin}i$ and $i$ denotes the inclination angle between the direction of the detector, as seen from the binary's center-of-mass, and the normal to the orbital plane which is assumed to be right-handed with respect to the sense of motion so that $0 \le i \le \pi$. $\delta m = m_1-m_2$, and the phase variable $\psi$ is given by
\begin{eqnarray}
\psi = \phi - 3 x^{3/2} \mbox{ln}\left(\frac{x}{x_0}\right),
\end{eqnarray}
where $\phi$ is the actual orbital phase of the binary, namely the angle oriented in the sense of motion between the ascending node and the direction of black hole 1 ($\phi = 0 ~ \mbox{mod}~2\pi$ when the two black holes lie along $\mathbf{u}$, with black hole 1 at the ascending node). The logarithmic phase modulation originates from the propagation of tails in the wave zone. The constant scale $x_0$ can be chosen arbitrarily; it relates to the arbitrary constant $b$ in the Eq. (\ref{eq219}). For details on higher order PN levels, see e.g. \cite{blanchet06}.

\begin{acknowledgement}
I thank the organizers of the Orl{\'e}ans School on Mass for their kind invitation and Luc Blanchet for helpful remarks on the manuscript.
\end{acknowledgement}

\end{document}